%% file: FabricEvolution.tex
\documentclass{jfm}
\usepackage{bm}
\usepackage{xcolor}
\usepackage{amsmath}
\usepackage{upgreek}
\usepackage{siunitx}

\input{command_defs}
\begin{document}

\shorttitle{Challenge to fabric evolution models} 
\shortauthor{R. N. Chacko et al.} 

\title{Shear Reversal in Dense Suspensions:\\
The Challenge to Fabric Evolution Models from Simulation Data}

\author{
 Rahul N. Chacko\aff{1},
  Romain Mari\aff{2},
  Suzanne M. Fielding\aff{1}
 \\
  \and
  Michael E. Cates\aff{2}
}

\affiliation{
\aff{1}
Department of Physics, Durham University, South Road, Durham DH1 3LE, UK
\aff{2}
DAMTP, Centre for Mathematical Sciences, University of Cambridge, Wilberforce Road, Cambridge CB3 0WA, UK
}

\maketitle

\input{abstract}

\input{intro} 
\input{simus} 
\input{standard_approach_section} 
\input{our_approach_section} 
\input{linear_and_quadratic} 
\input{2d_insight} 
\input{limitations} 
\input{discussion} 

\section*{Acknowledgements}
We would like to thank Morton M. Denn for pointing out the literature related to the Bingham closure.
The research leading to these results has received funding from
SOFI CDT, Durham University, and the EPSRC (grant ref. EP/L015536/1); and from
the European Research Council under the
European Union's Seventh Framework Programme (FP7/2007--2013) / ERC grant agreement number 279365.
MEC holds a Royal Society Research Professorship.

\appendix
\input{appendix_numerical_method} 
\input{appendix_componentwise_hand} 
\input{appendix_Qdot_decomp} 
\input{appendix_pdot_components_plot} 

\bibliographystyle{jfm}

\end{document}

%% file: command_defs.tex
\newcommand{\Tr}[1] {\mathrm{Tr}[#1]}
\providecommand{\abs}[1]{\lvert#1\rvert} 
\DeclareMathOperator{\sgn}{sgn}

\newcommand{\sg} {\dot{\gamma}}
\newcommand{\ag} {\abs{\sg}}
\newcommand{\sgg}{\sgn(\sg)}

\renewcommand{\vec}[1]{\bm{#1}}
\newcommand{\tsor}[1]{\bm{#1}}
\newcommand{\QQ} {\tsor{Q}}          

\newcommand{\dgam}[1]{{\dot{#1}}}
\newcommand{\dt}[1]{\frac{\mathrm{d}#1}{\mathrm{d}t}}
\newcommand{\dQ} {\dt{\QQ}}

\newcommand{\pSS}{\mathrm{+SS}}      
\newcommand{\mSS}{\mathrm{-SS}}      

\newcommand{\xmy}[1]{#1_{-}}
\newcommand{\xpy}[1]{#1_{+}}
\newcommand{\xy}[1]{#1_{12}}

\newcommand{\Id} {\tsor{I}}

\newcommand{\eq}[1]{(\ref{#1})}
\newcommand{\equ}[1]{Eq.~\eq{#1}}
\newcommand{\fig}[1]{(\ref{#1})}
\newcommand{\figu}[1]{Fig.~\fig{#1}}

%% file: abstract.tex

\begin{abstract}
Despite the importance of dense suspensions of hard particles as industrial
or environmental materials (fresh concrete, food, paint, or mud),
rather few constitutive models for them have been proposed or tested.
Most of these are, explicitly or effectively, ``fabric evolution models'' based on: (i) a stress rule
connecting the macroscopic stress to a second-rank microstructural fabric tensor $\QQ$;
and (ii) a closed time evolution equation for $\QQ$.
In dense suspensions most of the stress comes
from short-ranged pairwise steric or lubrication interactions at near-contacts (suitably defined),
so a natural choice for $\QQ$ is the deviatoric second moment of the distribution $P(\vec{p})$
of the near-contact orientations $\vec{p}$.
Here we test directly whether a closed time-evolution equation for such a $\QQ$
can exist, for the case of inertialess non-Brownian hard spheres in a Newtonian solvent.
We perform extensive numerical simulations accessing unprecedented levels of detail
for the evolution of $P(\vec{p})$ under shear reversal,
providing a stringent test for fabric evolution models.
We consider a generic class of these models as defined by~\citet{Hand1962}
that assumes little as to the micromechanical behaviour of the suspension
and is only constrained by frame indifference.
Motivated by the smallness of microstructural anisotropies in the dense regime,
we start with linear models in this class and successively consider those increasingly nonlinear in $\QQ$.
Based on these results we suggest that no closed fabric evolution model properly describes the dynamics of the fabric tensor under reversal.
We attribute this to the fact that, while a second-rank tensor captures reasonably well the microstructure in steady flows,
it gives a poor description during significant parts of the microstructural evolution following shear reversal.
Specifically, the truncation of $P(\vec{p})$ at second spherical harmonic (or second-rank tensor) level describes
ellipsoidal distributions of near-contact orientations, whereas on reversal we observe distributions that are markedly four-lobed;
moreover $\dot P(\vec{p})$ has oblique axes, not collinear with those of $\QQ$ in the shear plane.
This structure likely precludes any adequate closure at second-rank level.
Instead, our numerical data suggest that closures involving the coupled evolution of both a fabric tensor
and a fourth-rank tensor might be reasonably accurate.
\end{abstract}

%% file: intro.tex

\section{Introduction}

Non-Brownian suspensions of hard particles are commonly processed in many industries,
among them the ceramics, oil, construction, and food industries.
They usually contain particles of various shapes and sizes in
the range of a few microns. The suspended particles are large enough to experience no Brownian motion
and relatively weak interparticle interactions besides steric repulsion,
but are small enough for inertia to be neglected.
These suspensions have very simple physical ingredients,
yet they display a rather complex rheological behaviour.
They exhibit normal stress differences
~\citep{Zarraga2000,Singh2003,Couturier2011,Boyer2011a,Dai2013,Dbouk2013},
leading to unusual behaviours such as ``rod dipping''~\citep{Boyer2011a}.
Also they commonly exhibit shear thinning~\citep{gadalamaria_shearinduced_1980,Zarraga2000,Singh2003,Blanc2011a,Dai2013}
and, less commonly, shear thickening~\citep{Brown2009}.
For a recent review on these behaviours, see~\citet{Denn2014}.

Non-Brownian inertialess hard spheres,
density-matched to a Newtonian suspending fluid,
constitute the simplest idealized suspensions of this kind.
They show an apparently simple rate-independent rheology, albeit a non-Newtonian one.
They have finite normal stress differences, $N_2$ being negative and dominant over $N_1$,
whose sign is still debated~\citep{Denn2014}.
Furthermore, there is a strong dependence of the rheology on the deformation history,
as exemplified by the complex response observed in shear reversal experiments~\citep{gadalamaria_shearinduced_1980}
or time-dependent deformations~\citep{blanc_tunable_2014}.
Rate independence follows from the fact that hard-sphere interactions have no characteristic stress scale
or time scale. This means that the microstructure depends on strain history but not strain-rate history.
Accordingly the microstructure in steady flow is independent of the strain rate. All stress components
are then linear in strain rate~\citep{Denn2014}. Yet, despite these simplifications, and despite the practical importance
of predictive models for dense suspension mechanics,
rather few constitutive equations have been proposed
~\citep{Hinch1975,Phan-Thien1995,Stickel2006,Goddard2006}.

Constitutive models of rheology have been successfully developed
in the context of polymer solutions and melts~\citep{larson2013constitutive}.
The simplest models (e.g. Johnson-Segalman~\citep{johnson_model_1977} or Giesekus~\citep{giesekus_simple_1982} models)
are based on a time-evolution equation of the stress tensor
involving the stress tensor itself as well as the strain-rate tensor.
While these models can be interpreted as consisting of two separate pieces,
a proportionality relation between the macroscopic stress and the molecular conformation tensor,
and a closed dynamical equation for the conformation tensor,
the combination of these two elements leads to a single closed dynamical equation
for the stress tensor, and the fabric evolution is implicit.
However, by construction these models have a finite relaxation time and can only predict
a continuous time-evolution for the stress, a feature incompatible with the discontinuities observed
in non-Brownian suspensions under discontinuous changes of imposed flow like shear reversal~\citep{gadalamaria_shearinduced_1980},
due to the one-sidedness of contact forces.
Instead, for non-Brownian suspensions the constitutive models may require one to explicitely keep
distinct the recipe linking the macroscopic stress to one or more microstructural state variables
(which could then be explicitly alert to discontinuities in the imposed flow)
and the closed (and continuous) dynamics for the microstructure.
These two separate problems can be formulated top-down in a purely phenomenological manner,
or derived bottom-up from the
many-body microscopic dynamics, through either a coarse-graining procedure or a mechanistic approximation.
In all cases, the resulting model is constrained by symmetry
and frame indifference considerations~\citep{Hand1962}.
The constitutive models for non-Brownian suspensions of~\citet{Hinch1975,Phan-Thien1995,Stickel2006,Goddard2006}
follow this path, and pick for the microstructural variable a second-rank tensor
$\QQ$ called fabric tensor.
A second-rank fabric tensor is the simplest object capturing anisotropies
in the pair interaction network
and is a natural candidate to inform a stress equation.
Indeed, it has been recognized for some time that microstructural
anisotropies play a central role in the stress response of dense suspensions
~\citep{Batchelor1972,gadalamaria_shearinduced_1980,Wagner1992,blanc_microstructure_2013,Gurnon2015}.
Perhaps surprisingly, while they are apparently well-founded
and follow conceptually from a history of successful polymer constitutive modelling,
none of the fabric evolution equations for non-Brownian dense suspensions proposed so far
has been thoroughly tested by comparison with detailed experimental or numerical data.

There are experimental reasons for this.
While rheometric measurements for dense suspensions have been obtained
under many different flow geometries and conditions
(rate controlled, stress controlled
or even particle-pressure controlled~\citep{Boyer2011}),
it is still challenging to obtain
finely resolved microstructural data under rheometric flow.
This is currently only accessible through confocal microscopy
owing to the particle sizes involved~\citep{cheng_imaging_2011}, and even then, the presence of even mild polydispersity
confounds attempts to interrogate accurately the statistics of near-contacts.
Note that the problem is spatial and not temporal resolution: in the rate-independent case
flow can be stopped at any time to inspect the structure and restarted without changing the subsequent dynamics.

Numerical simulations, on the other hand,
can provide a high level of detail in the structure,
and probe different flow conditions.
Simple shear is by far the most common flow geometry studied,
but others are possible, including planar elongational flow~\citep{hwang_direct_2006,seto_microstructure_2017}.
Numerical data can be used to test the validity of constitutive models,
but also to guide the development of new ones
once relevant physical microscopic mechanisms are identified.
The use of simulations is however
subject to prior validation of the physical assumptions
on which the numerical method is based,
and this has proved a challenge for dense suspensions.
The last few years have seen a breakthrough in this area with
the recognition of the role of frictional contacts in the microstructure dynamics
and stress response~\citep{Brown2012,Seto2013,Mari2014,Guy2015,Lin2015a},
based on earlier experimental evidence
~\citep{castle_effect_1996,Lootens2005,Blanc2011}.
This has led to the development of numerical simulations which for the first
time quantitatively match the experimental data
for the rheology of dense suspensions~\citep{Gallier2014,Mari2015,Lin2015a}.
We recall that particle friction can preserve the rate independence of the rheology,
provided that it does not introduce a rate or stress scale;
many simple models of friction, like Coulomb friction used in this work, indeed lead to
a rate-independent rheology~\citep{Mari2014,Gallier2014}.

In this paper we leverage these new numerical simulation capabilities
to interrogate the validity
of the basic assumptions underlying fabric evolution models of the type
so far used to build phenomenological constitutive equations.
We present detailed numerical results for Stokesian dense
suspensions of hard spheres interacting solely through near-field lubrication and
Coulomb frictional contacts and subject to a shear reversal protocol.
Shear reversal is an informative probe of rheological mechanisms,
and various experiments have measured the stress response under reversal
for non-Brownian suspensions~\citep{gadalamaria_shearinduced_1980,Kolli2002,Narumi2002,Blanc2011a}.
For hard-sphere suspensions,
the shear reversal protocol
retains the simplicity of shear flow
while being one of the most stringent test cases
for a microstructural description,
as it probes the dependence on deformation history
in perhaps its most extreme realization.
Indeed, upon reversal, both hydrodynamic
and contact force amplitudes are discontinuous.
The hydrodynamic forces follow the discontinuous change of
velocity field in the solvent, and a finite fraction of contacts
are put under tension and thus instantaneously open,
forcing a global load rebalance on the contact force network.
Moreover, shear reversal is known to probe separately
lubrication- and friction-dominated strain regimes~\citep{Lin2015a,Ness2016,peters_rheology_2016}.

We find below that quantitative agreement between the fabric tensor dynamics
and simulation data cannot be achieved in fabric evolution models;
at least not those that lend themselves to a simple physical interpretation.
By interrogating the simulation data,
we explore why fabric evolution in these systems
is not well described by closed equations
involving only the fabric tensor and the imposed flow.
As will become clear, the fabric dynamics involves more details of the microstructure,
some of which may be captured via higher spherical harmonics
of the angular distribution of near-contacts.

In section~\ref{sec:simus}, we explain our simulation protocol for aquiring
shear reversal data. In section~\ref{sec:standard_hand}, we overview the fabric evolution approach, starting from the results of~\citet{Hand1962}
for the frame-invariant dynamics of a second-rank tensor.
In section~\ref{sec:our_hand}, we introduce the family of fabric tensors
whose dynamics we will consider, and start to
develop models of increasing complexity within the Hand framework.
In section~\ref{sec:results}, we show that the simplest families
of possible fabric dynamics, restricted to linear or quadratic closures, fail to describe
the dynamics of the system upon reversal. Using insights from two-dimensional
simulations, in section~\ref{sec:2d_insight} we show that models based on higher order closures
may fit the numerical data but are unlikely to be physically informed,
and instead owe their fitting to the number of free parameters they contain.
In section~\ref{sec:discussion}, we argue that the failure
of fabric evolution models stems not from truncation at finite order in $\QQ$ but from the fact that the full distribution of near-contact directions $P(\vec{p})$
is dominated by a fourth-rank component, not coaxial with $\QQ$, during part of the evolution following shear reversal.
However we speculate that the evolution of the microstructure dynamics
might be satisfactorily closed
by including both a second-rank fabric tensor $\QQ$ and a fourth-rank tensor that is not a function of the fabric and/or strain-rate tensors.

%% file: simus.tex

\section{Simulation protocol}\label{sec:simus}

We simulate an assembly of non-inertial frictional spheres
immersed in a Newtonian fluid under simple shear flow
with an imposed velocity field
$\vec{v} = (\sg y, 0, 0)$,
using Lees-Edwards periodic boundary conditions~\citep{Lees1972}.
The system is binary, with radii $a$ and $1.4a$ mixed at equal volume fractions.
The particles interact solely through frictional contacts (with a friction coefficient $\mu=1$)
and short-range hydrodynamic forces (lubrication) with a resistance divergence at contact truncated
by a typical roughness length scale $\delta=10^{-3}a$~\citep{Seto2013,Ness2016}.
The contacts are modeled with the Cundall-Strack model~\citep{cundall_discrete_1979}
and thus are slightly deformable for purely numerical reasons.
We use particles' stiffness such that the overlaps are kept below a maximum of \SI{2}{\percent} of
a particle radius, which ensures a hard-sphere behavior.
We have tested that a doubling of the maximal overlap value
does not affect the dynamics of the fabric tensor past a strain of
roughly \SI{1}{\percent} after reversal during which there is an elastic recoil
from the contacts.

With these effectively hard-sphere interactions, the value of $\sg$ only sets the speed at which the particles move
on otherwise rate-independent trajectories.
We performed the simulations at three volume fractions $\phi=0.4, 0.5$ and $0.55$ with $N=500$ particles.
For reference, the jamming point for this system is at $\phi_{\mathrm{J}} \approx 0.58$~\citep{Mari2014}.
The simulation method, which is detailed in~\cite{Mari2014},
is briefly outlined in appendix~\ref{sec:appendix_numerics}.

Starting from an overlap-free random initial configuration,
the system is sheared in the $+x$ direction for 5 strain units, by which point
the system has reached steady state.
We then perform a flow reversal, and shear the system in the $-x$ direction
for 3 strain units, during which structure is monitored continuously.
In order to improve the statistics of our results, we repeat this procedure from
250 steady state configurations.
The data presented here are obtained from an average of these 250 flow reversal realizations.

%% file: standard_approach_section.tex

\section{Fabric-based microstructure description}\label{sec:standard_hand}

\subsection{Fabric tensor}

In a non-Brownian hard-sphere suspension,
particles interact solely through hydrodynamic and hard contact forces.
While contact forces are short range pairwise interactions, hydrodynamic forces result
from the superposition of algebraically decaying solvent velocity fields created by the particles' motion,
and as such have a long-range and many-body nature.
In a concentrated system, however, most of the hydrodynamic tractions on the particle surfaces
come from lubrication flows within the narrow interparticle gaps~\citep{Frankel1967}.
These lubrication forces are pairwise and short-ranged in nature.
Therefore, in the dense regime, most of the stress comes from pairwise short-range interactions~\cite{ball_simulation_1997,Mari2015};
indeed the long-range hydrodynamics is omitted from our simulations as described above.
The pairwise nature of dominant forces will give a special importance for stress prediction to the pair correlation function
$g(\vec{r})$, defined as the conditional probability to find a particle whose center is at $\vec{r}$
knowing that there is a particle at $\vec{0}$, normalized by the average number density $\rho$.
Moreover, because the dominant forces are also short-ranged and $g(\vec{r})$
is peaked at or very near contact~\citep{morris_microstructure_2002,nazockdast_microstructural_2012,nazockdast_pair-particle_2013,Mari2014},
we can expect that only the near-contact part of $g(\vec{r})$ will play a substantial role.
Hence we can consider the distribution of near-contact orientations $P(\vec{p})$ (with $\vec{p} \in S^2$),
which keeps only the orientational information of interactions which determine
the stress state of the suspension.

Building a time evolution dynamics for $P(\vec{p})$ (let alone $g(\vec{r})$)
is however very challenging for a non-Brownian suspension,
and quite possibly over-reaching in the context of building a constitutive model.
Indeed, the stress $\tsor{\Sigma}$ is a symmetric second-rank tensor with $6$ independent components,
and in consequence contains far less information than the entire $P(\vec{p})$.
Therefore, rather than attempt to model the full $P(\vec{p})$
(or $g(\vec{r})$~\citep{nazockdast_microstructural_2012,nazockdast_pair-particle_2013}) directly,
previous attempts model the symmetric and traceless fabric tensor
$\QQ := \langle \vec{p} \vec{p} \rangle - (1/3) \Id$.
This is the de-traced second moment of $P(\vec{p})$, and the lowest order moment
carrying nontrivial structural information.
It also corresponds to the second-order term in the Laplace spherical harmonic expansion of $P(\vec{p})$
in tensor form~\citep{ken-ichi_distribution_1984}\footnote{For a symmetric distribution on the unit sphere like $P(\vec{p})$, it is possible to define a tensorial expansion
in powers of $\vec{p}$ as $P(\vec{p}) = \frac{1}{4\pi}\sum_{s=0}^{\infty} D_{i_1,\ldots, i_{2s}} p_{i_1}\cdots p_{i_{2s}}$
such that the term of order $n=2s$ in the tensorial expansion is the projection of $P(\vec{p})$ on the $2n+1$-dimensional space
of Laplace spherical harmonics of order $n$~\citep{ken-ichi_distribution_1984}.
Put differently, the order $n$ term of the tensorial expansion is the sum of
the terms of order $n$ in the Laplace spherical harmonics expansion.
The tensors $D_{i_1,\ldots, i_{n}}$ are deviatoric, hence contain $2n+1$ independent components,
and are mutually independent, a consequence of the orthogonality of the Laplace spherical harmonics basis.
The first term tensor ($0$th rank) of the tensorial expansion is $1$
(which enforces the normalization of $P(\vec{p})$), the tensor of the second term is $\frac{15}{2} \QQ$.},
\begin{equation}
P(\vec{p}) \approx \frac{1}{4 \pi} \left( 1 + \frac{15}{2} \QQ:\vec{p}\vec{p}\right). \label{she2}
\end{equation}

Retaining the fabric tensor $\QQ$ as a proxy for the full microstructure
is indeed the standard choice made in previous attempts to build a constitutive model
of suspensions; see~\citet{Hinch1975,Phan-Thien1995,phan-thien_new_1999,Stickel2006,Goddard2006}.
Note that there is also a long history of representing the microstructure
via a second-rank tensor in polymeric systems~\citep{larson2013constitutive}
and in dry granular materials~\citep{Sun2011,magnanimo_local_2011,goddard_continuum_2014}.

\subsection{Dynamics: Hand equation}

Once one decides to write the dynamics of the microstructure as
a closed set of ODEs for the time evolution of the fabric tensor,
symmetries and frame indifference (also known as material objectivity)
constrain quite strongly the functional forms involved.
In the absence of
inertia, systems are invariant under time-dependent translations and rigid
rotations, a property known as frame indifference.
Using a representation theorem and enforcing frame-indifference,
one can write a general evolution equation for a symmetric $3\times3$ tensor~\citep{Hand1962}. Specifically if
$\dQ$ depends only on $\QQ$ itself and on the rate-of-strain tensor $\tsor{K}$, we have:
\begin{multline}
\dQ
= \tsor{W} \cdot \QQ - \QQ \cdot \tsor{W}
+ \alpha_0 \Id
+ \alpha_1 \QQ
+ \alpha_2 \tsor{E}
+ \alpha_3 \QQ^2
+ \alpha_4 \tsor{E}^2 \\
+ \alpha_5 \left( \tsor{E}\cdot \QQ + \QQ\cdot \tsor{E} \right)
+ \alpha_6 \left( \tsor{E} \cdot \QQ^2 + \QQ^2 \cdot \tsor{E} \right)\\
+ \alpha_7 \left( \tsor{E}^2 \cdot \QQ + \QQ \cdot \tsor{E}^2 \right)
+ \alpha_8 \left( \tsor{E}^2 \cdot \QQ^2 + \QQ^2 \cdot \tsor{E}^2 \right),~\label{eq:Handeq_general}
\end{multline}
where $\tsor{W}$ and $\tsor{E}$ are respectively the
antisymmetric and symmetric parts of $\tsor{K}$,
and where the scalar coefficients
$\alpha_i$ are analytic functions of the following invariants
$ I_1 := \Tr{\QQ}$,
$ I_2 := \Tr{\QQ^2}$,
$ I_3 := \Tr{\QQ^3}$,
$ I_4 := \Tr{\tsor{E}}$,
$ I_5 := \Tr{\tsor{E}^2}$,
$ I_6 := \Tr{\tsor{E}^3}$,
$ I_7 := \Tr{\QQ \tsor{E}}$,
$ I_8 := \Tr{\QQ^2 \tsor{E}}$,
$ I_9 := \Tr{\QQ \tsor{E}^2}$,
and
$ I_{10} := \Tr{\QQ^2 \tsor{E}^2}$.
Note that \equ{eq:Handeq_general} follows only from frame indifference
and the fact that $\QQ$ is a symmetric 3-by-3 tensor.
It does not rely on any specific physical property of the system nor
on an expansion to some order in $\QQ$ or $\tsor{E}$.
As such, \emph{any} closed ODE for $\dQ$ analytic in $\QQ$ and $\tsor{K}$
has to be of the form of \equ{eq:Handeq_general}.

In our case, \equ{eq:Handeq_general} is further constrained.
First, $\QQ$ is traceless which implies that $I_1=0$
and $\alpha_0$ is prescribed by the value of the other coefficients $\alpha_i$.
Furthermore, for a rate-independent system the right-hand side of~\equ{eq:Handeq_general}
must be proportional to the absolute value of the strain rate $\ag=\sqrt{2\tsor{E}:\tsor{E}}$.
Also, at least for the simple shear flows studied below, we can change variable from time to the accumulated strain $\gamma$.
Hence, by introducing $\tsor{\hat{E}}=\tsor{E}/\ag$, $\tsor{\hat{W}}=\tsor{W}/\ag$
and $\dgam{\QQ} = \mathrm{d}\QQ/\mathrm{d}\gamma$, we can rewrite the Hand equation as
\begin{multline}
  \sgg\dgam{\QQ}
  = \tsor{\hat{W}} \cdot \QQ - \QQ \cdot \tsor{\hat{W}}
  + \alpha_0 \Id
  + \alpha_1 \QQ
  + \alpha_2 \tsor{\hat{E}}
  + \alpha_3 \QQ^2
  + \alpha_4 \tsor{\hat{E}}^2\\
  + \alpha_5 \left( \tsor{\hat{E}} \cdot \QQ + \QQ \cdot \tsor{\hat{E}} \right)
  + \alpha_6 \left( \tsor{\hat{E}} \cdot \QQ^2 + \QQ^2 \cdot \tsor{\hat{E}} \right) \\
  + \alpha_7 \left( \tsor{\hat{E}}^2 \cdot \QQ + \QQ \cdot \tsor{\hat{E}}^2 \right)
  + \alpha_8 \left( \tsor{\hat{E}}^2 \cdot \QQ^2 + \QQ^2 \cdot \tsor{\hat{E}}^2 \right).~\label{eq:Handeq_rate_indep}
\end{multline}
where the $\alpha_i$'s are now functions of the 9 invariants of $\hat{I}_k$ involving $\QQ$ and $\tsor{\hat{E}}$.

Hand's equation in the form~\eq{eq:Handeq_rate_indep}
provides the boundaries within which one can build a dynamics
for $\QQ$. It is nonetheless not practically useful as it stands,
as there is no prescription for the coefficients $\alpha_i$,
and the space of allowed dynamical equations is still infinite dimensional.
At this point, one usually has to introduce further asumptions about
the specific nature of the dynamics in order
to reduce the number of free parameters to a handful or so,
based on physical motivations or simply tractability of the model.

%% file: our_approach_section.tex

\section{Systematics of Hand-based models}\label{sec:our_hand}

\subsection{A family of fabric tensors}

There is some freedom in the exact definition of $\QQ$.
We would like $\QQ$ to be an average over the interactions that are relevant
to the suspension stress, which we know are short-range.
What do we mean by short-range interaction precisely?
One could consider a $\QQ$ including only strict contact interactions, or including every pair of particles sharing an edge in a Delaunay tesselation
(i.e. particles occupying neighboring cells in a Voronoi tesselation),
or including every pair of particles separated by a gap smaller than $\epsilon$.
We adopt the latter choice and define the set of directions of interactions closer than a gap $\epsilon$ as
$\Gamma^{\epsilon} = \{\vec{p}_{ij}\  \forall i, j \mid 2(r_{ij}/(a_i+a_j)-1)<\epsilon\}$, where $i,j$ are particle labels, $r\vec{p}$ is a centre-to-centre vector and $a$ is a particle radius.
We then obtain a family of traceless fabric tensors $\QQ^{\epsilon}$ as
\begin{equation}
  \QQ^{\epsilon} = \langle \vec{p} \vec{p} \rangle_{\vec{p} \in \Gamma^{\epsilon}} - (1/3) \Id .
\end{equation}

\begin{figure}
  \centering
  \includegraphics[width=0.48\textwidth]{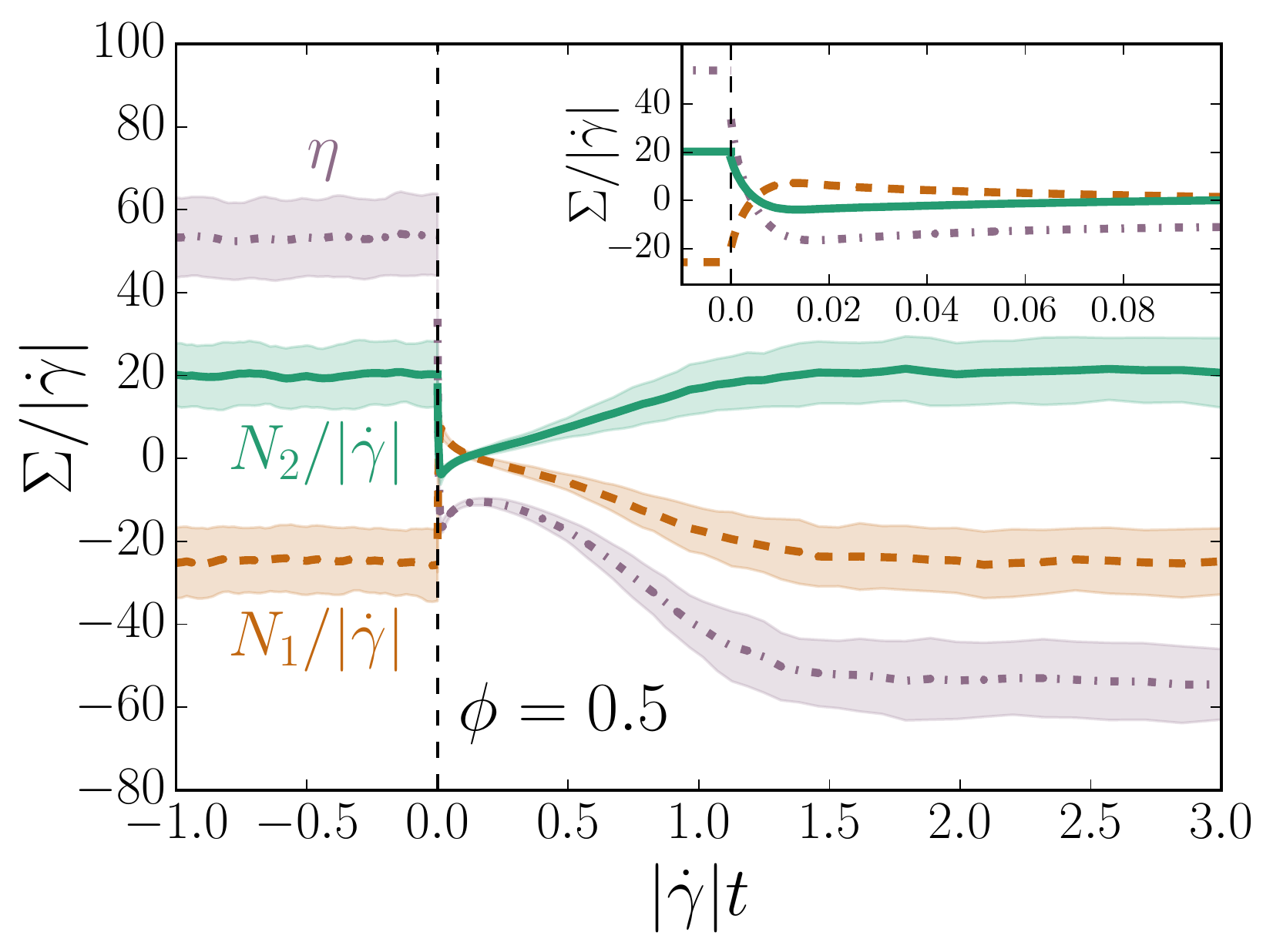}
  \includegraphics[width=0.48\textwidth]{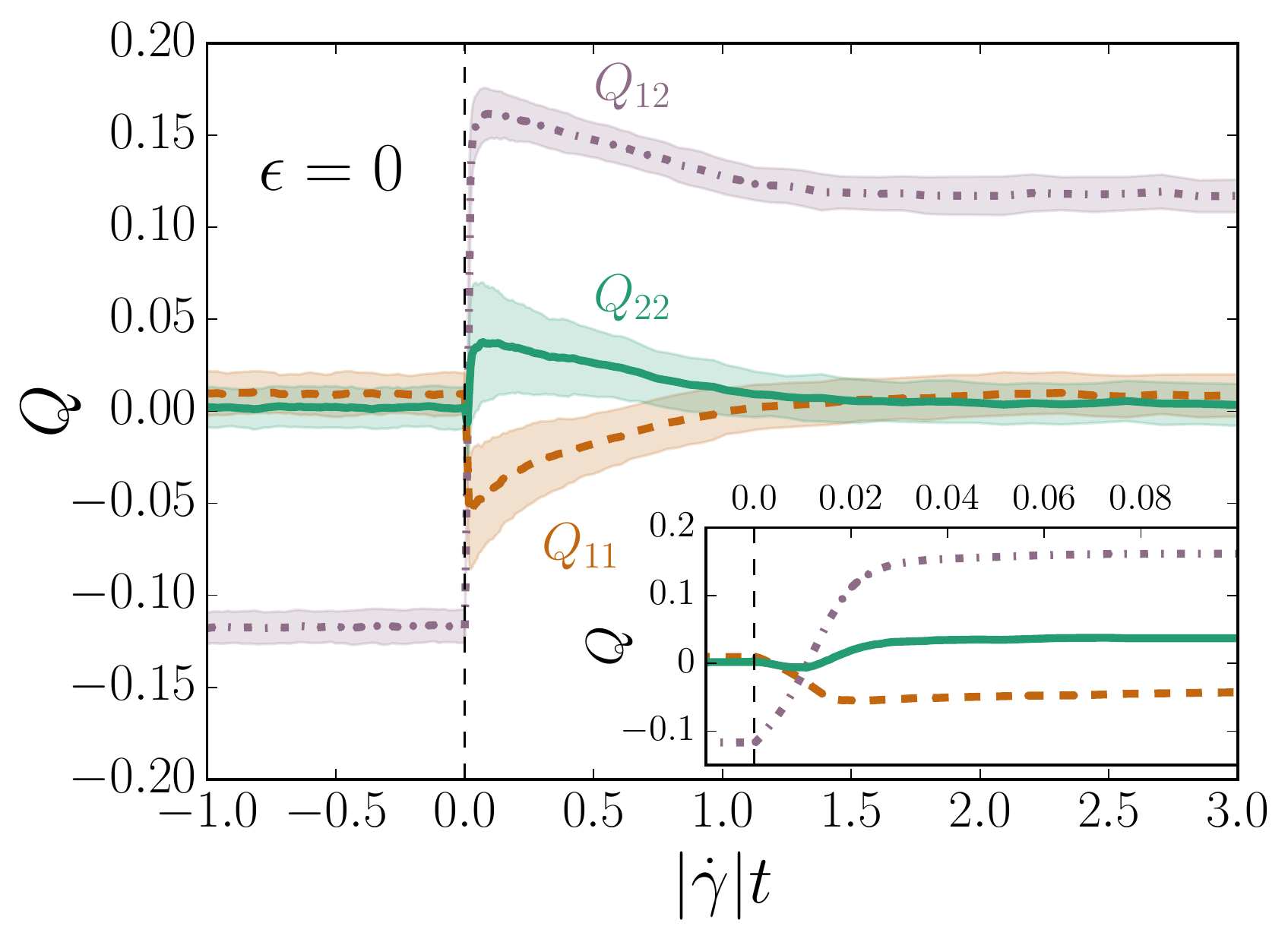}
  \caption{Left: Stress tensor components upon shear reversal at $\gamma=0$ for $\phi=0.5$.
  $\Sigma_{13}$ and $\Sigma_{23}$ are not shown as they vanish in simple shear by symmetry.
  Thick dark-shaded lines are the averaged data, while the light shaded area around each curve
  is the standard deviation obtained from the individual shear reversals.
  All components show a discontinuity upon reversal, arising from lubrication and/or contact forces.
  Right: Fabric tensor components (with $\epsilon = 0$, describing full contacts only) for the same conditions, with averages in thick lines and standard deviation in shaded areas.
Not shown are $Q_{13} = Q_{23} = 0$ and  $Q_{33} = -Q_{11}-Q_{22}$.}
\label{fig:S_numerics}
\end{figure}

\begin{figure}
  \centering
  \includegraphics[width=0.99\textwidth]{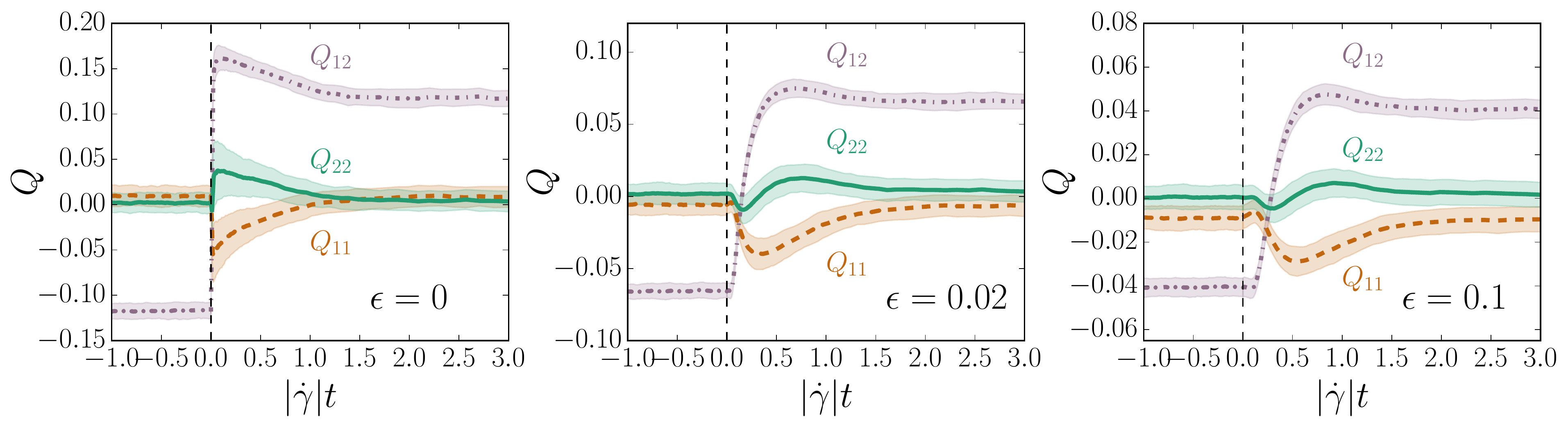}
  \caption{Fabric tensor components upon reversal at $\gamma=0$ for $\phi=0.5$
  and for 3 different coarse-graining $\epsilon=0, 0.02, 0.1$.
  The fabric tensor built on contacts (that is, with $\epsilon=0$) is discontinuous upon reversal,
  as a finite fraction of contacts disappear when the flow is reversed.
  However, when the coarse-graining length is finite, the discontinuity disappears,
  and instead the fastest characteristic evolution
  seems to take place over a strain scale of order $\epsilon$.}
\label{fig:Q_numerics_delta_dep}
\end{figure}

Using numerical simulations,
we can study the dynamics of $\QQ^{\epsilon}$ for any $\epsilon$.
In \figu{fig:S_numerics}, we show that the dynamics of $\QQ^{\epsilon=0}$
(i.e., including only strict contacts, which correspond to interactions
with a small negative gap thanks to the finite deformability of the particles)
and $\tsor{\Sigma}$
share several features upon reversal.
In particular they both show a discontinuity exactly at reversal,
followed by a relaxation back to steady state over the same strain scale,
here roughly 2 strain units for $\phi=0.5$.
(Strictly speaking, because of the slight deformability of the particles, the apparent
discontinuity in $\QQ^{\epsilon=0}$ at reversal is actually a fast relaxation taking place
during the first \SI{1}{\percent} of strain after reversal.)
Because of the symmetry of the simple shear flow,
both admit nonzero values only for their diagonal components
and for their shear components in the velocity-gradient plane, $Q_{12}$ and $\Sigma_{12}$.
The steady-state values of these components have the same parity under strain reversal,
with sign reversal for shear and none for the diagonal elements, providing further motivation for the use of $\QQ$
as a proxy for the microstructure in a constitutive model for the stress $\tsor{\Sigma}$.

However, despite the appeal of the similarity with the dynamics of the stress,
a $\QQ$ discontinuous upon reversal does not lie in the realm of ODEs
within the Hand family defined in \equ{eq:Handeq_rate_indep}.
Indeed, in \equ{eq:Handeq_rate_indep}, $\dgam{\QQ}$ remains finite
at reversal, irrespective of the values of the $\alpha_i$'s,
and we conclude that with such a dynamics $\QQ$ cannot show a discontinuity.
In \figu{fig:Q_numerics_delta_dep}, we show instead that for $\epsilon>0$,
$\QQ^{\epsilon}$ is continuous upon reversal,
and otherwise keeps most of the qualitative features of $\QQ^{\epsilon=0}$.
The initial singular reversal behavior is smoothed
over a finite strain scale which is increasing with $\epsilon$ and saturates
for large $\epsilon$ to a strain $O(1)$.
Separately, the overall amplitude of $\QQ^{\epsilon}$ decreases with $\epsilon$.

An appealing strategy is then to assume that knowledge of the fabric of near-contacts, $\QQ^{\epsilon>0}$,
along with knowledge of the flow itself,
is enough to predict the evolution of the stress tensor $\tsor{\Sigma}$.
That is, a continuous time evolution of $\QQ$ obeying an ODE in the Hand class might be married
with a rule for constructing $\tsor{\Sigma}(\QQ, \tsor{K})$
that is explicitly alert to discontinuities at the moment of strain reversal.
(This rule could build in information about the sudden opening of contacts on scales below $\epsilon$, for instance.)
An alternative strategy is to build a time evolution for the true contact fabric ($\epsilon=0$) based on two relaxation times,
one large corresponding to the time scale of reorganization of the microstructure, and one extremely small
to describe the instantaneous changes in the contacts upon flow changes~\citep{Goddard2006}.
However, with this approach the fit to actual data imposes the parameters associated
with the short time relaxation to take values whose physical interpretation is complex,
including for instance transiently negative diffusion coefficients~\citep{Goddard2008}.

We stress that the coarse-graining length $\epsilon$ is not related to the lubrication regularization length $\delta$
nor to the maximum allowed overlap in the contact model.
Moreover, it appears that because $\QQ^{\epsilon}$ is getting smoothed out with increasing $\epsilon$,
the effects of contact overlaps and $\delta$, which are only visible in particle trajectories at very early strains after reversal,
will be gradually hidden in the time evolution of $\QQ^{\epsilon}$ for increasing $\epsilon$.
In practice, we find that $\QQ^{\epsilon}$ becomes insensitive to a doubling of $\delta$ for $\epsilon$ as small as $0.01$.
With this as the modelling strategy, picking the  best coarse-graining length ${\epsilon}$ to build a consitutive model is then
a compromise between the accuracy needed to admit a quantitative stress rule from $\QQ^{\epsilon}$,
and the smoothness required for an ODE closure in the $\QQ$ dynamics.
In this work, we do not address the issue of the stress rule further,
addressing solely the dynamics of $\QQ$ itself.
We will focus our numerical tests of fabric evolution models on the case
$\epsilon=0.02$, adopting the default notation $\QQ:=\QQ^{\epsilon=0.02}$.
But, as shown in Figs.~\fig{fig:Q_numerics_delta_dep}~and~\fig{fig:Q_delta_phi_dep}, the qualitative features of the dynamics
are shared across all values of $\epsilon$, so that most of what we will learn by trying to fit the case $\epsilon=0.02$
readily extends to other values of $\epsilon$.

\subsection{Weak anisotropy: Small $\QQ$ models}

\begin{figure}
  \centering
  \includegraphics[width=0.99\textwidth]{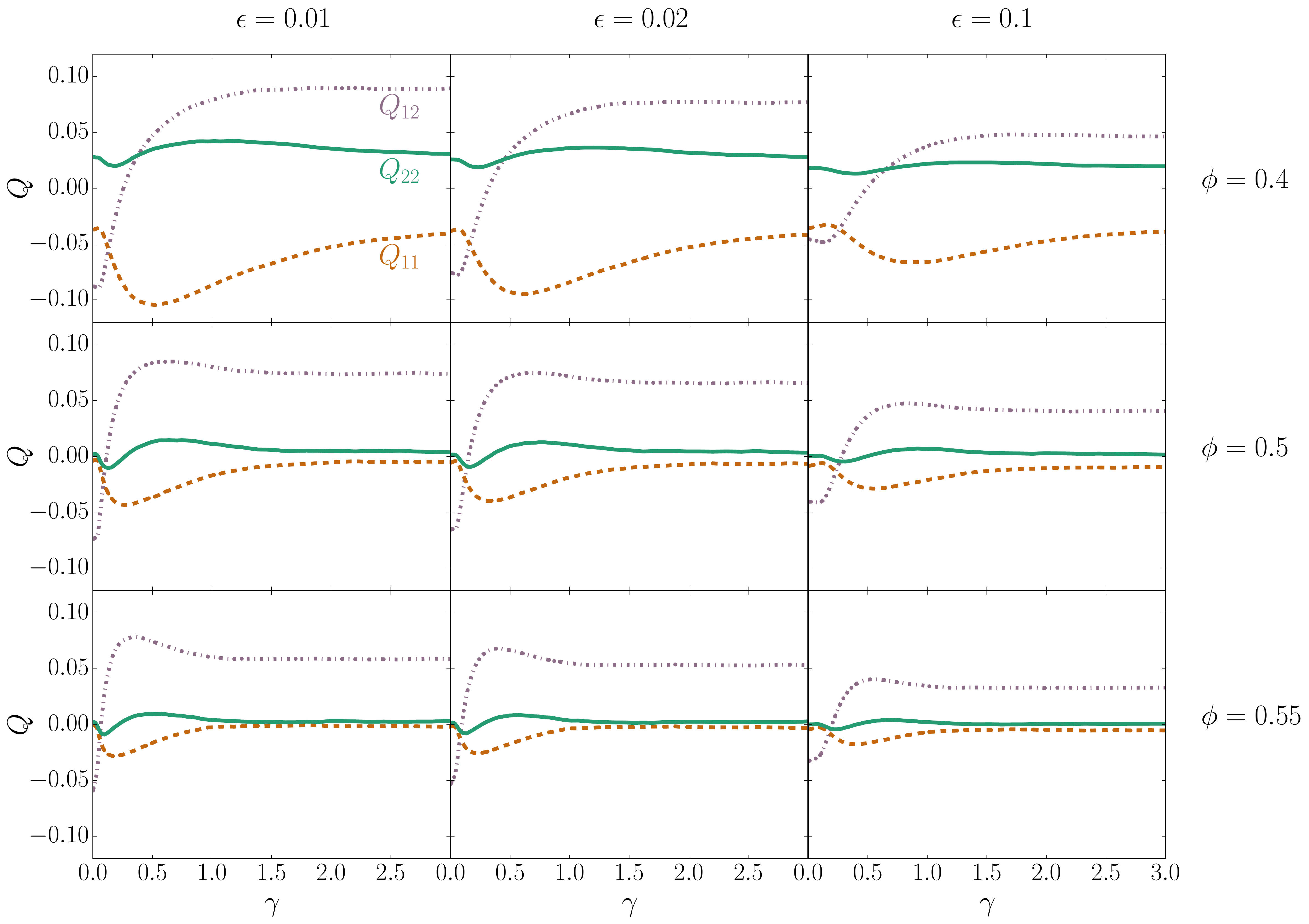}
  \caption{Fabric tensor components upon reversal at $\gamma=0$
  for 3 volume fractions $\phi=0.4,0.5, 0.55$ (respectively first, second and third row)
  and for 3 different coarse-graining scales $\epsilon=0.01, 0.02, 0.1$ (respectively first, second and third column).}
\label{fig:Q_delta_phi_dep}
\end{figure}

At large volume fractions, the steric constraints impose that every particle in the suspension
is surrounded by close neighbors in any direction,
with possibly only one direction (the compressional axis in simple shear)
being slighly more crowded than the others.
We can then expect that the anisotropy of the microstructure is decreasing
with increasing $\phi$.

\begin{figure}
  \centerline{\includegraphics[width=0.5\textwidth]{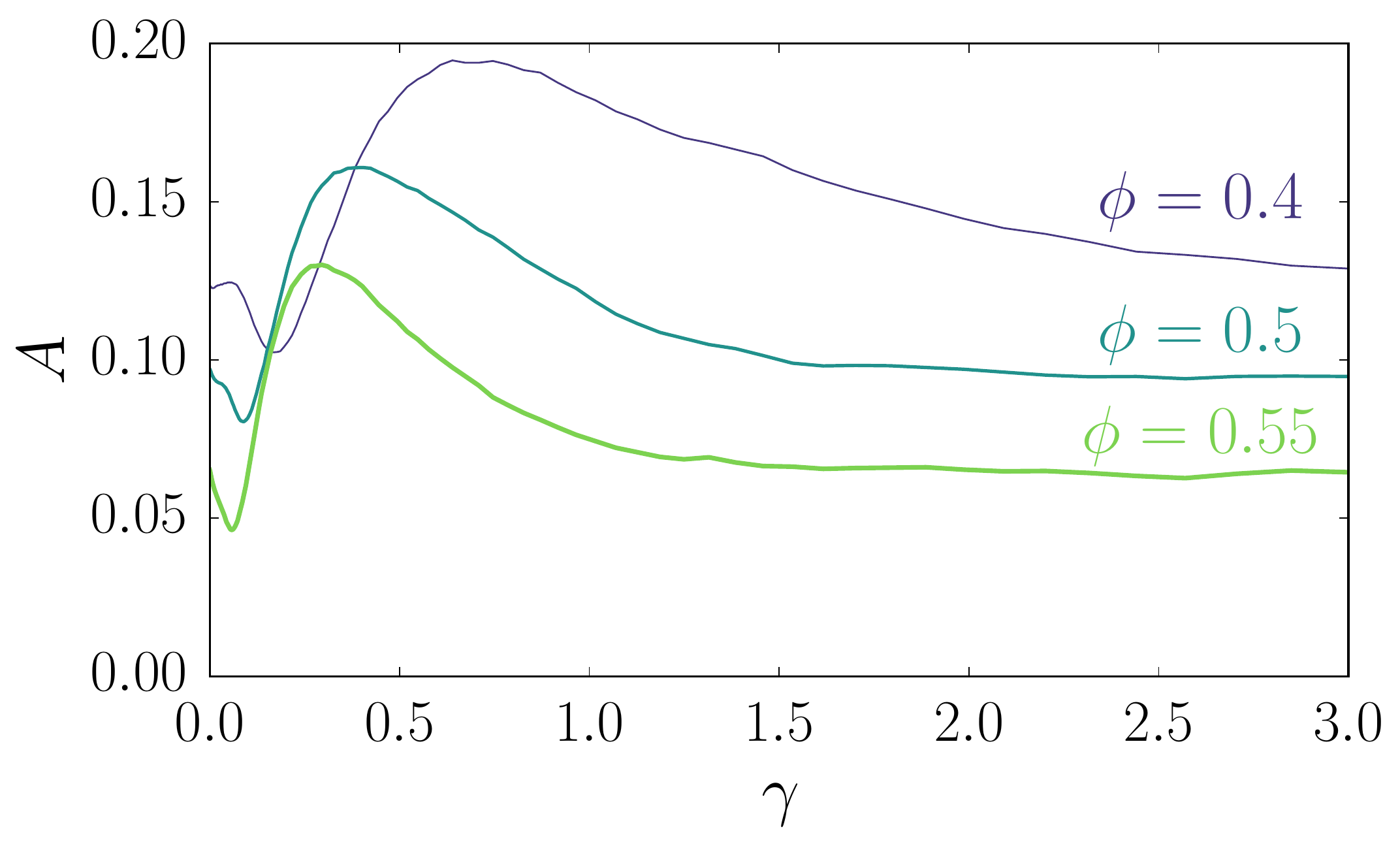}}
  \caption{Anisotropy $A$ as a function of post-reversal strain for three volume fractions
  $\phi=0.4, 0.5$ and $0.55$.
\label{anifig}}
\end{figure}

As evidence for this, given a distribution of near-contacts $P(\vec{p})$,
we quantify its anisotropy by defining
$A := \left(4\pi\right)^{-1}\int_{S^2} \left[ 4\pi P(\vec{p}) - 1 \right]^2 \mathrm{d}\Omega$.
\figu{anifig} shows the post-reversal evolution of $A$
for various volume fractions $\phi$,
and makes it clear that $A$ does indeed decrease with increasing $\phi$.
Similarly, in \figu{fig:Q_delta_phi_dep}, we show the dependence of $\QQ$
on $\phi$, for several values of $\epsilon$, which demonstrates for $\QQ$ the
corresponding lowering of the anisotropy as $\phi$ increases.
To quantify further how small $\QQ$ actually is,
we can look at a norm such as $\left|\QQ\right| = \sqrt{\Tr{\QQ^2}}$
.
It is easy to show that $\left|\QQ\right| \leq \sqrt{2/3} \approx 0.816$,
with equality corresponding to having
all the $\vec{p}$ in the same direction, $P(\vec{p})=\delta(\vec{p}-\vec{p}_0)$.
In contrast with this saturation value, we observe in steady state for $\phi=0.4$ and $\epsilon=0.01$
(the top left case in \figu{fig:Q_delta_phi_dep}),
$\left|\QQ\right|\approx 0.048$, and for $\phi=0.55$ and $\epsilon=0.01$
(bottom left), $\left|\QQ\right|\approx 0.0022$. These figures, which confirm a weak anisotropy limit, become even smaller if we increase $\epsilon$.
A reasonable assumption, based on the smallness of these values, is to restrict
the evolution equation~\eq{eq:Handeq_rate_indep} to low orders in $\QQ$,
allowing a Taylor expansion of the dynamics around the isotropic state.
In what follows, we will examine the resulting dynamical equations
at successively higher orders in $\QQ$.

\subsection{Hand equation in simple shear}

From now on, we restrict our discussion to a simple shear flow
with $\hat{K}_{ij} = 0$ except $\hat{K}_{12}=\pm 1/2$,
which is the simulation case.
Because of the sparsity of $\tsor{\hat{K}}$ in this geometry,
not all the components of the right-hand side of \equ{eq:Handeq_rate_indep}
are linearly independent, and some of the $\alpha_i$'s are redundant.
Hence we recast the tensorial Hand equation for rate-independent systems \eq{eq:Handeq_rate_indep}
in a non-redundant component form (see Appendix~\ref{Handapp}) as
\begin{equation}
  \begin{aligned}
  \xmy{\dgam{Q}}
  &= \xmy{P}\left[ \xpy{Q},\sgg \xy{Q}, \xmy{Q}^2 \right] \sgg \, \xmy{Q}
  + 2 \xy{Q} \\
  \xpy{\dgam{Q}}
  &= \xpy{P}\left[ \xpy{Q},\sgg \xy{Q}, \xmy{Q}^2 \right] \sgg \\
  \xy{\dgam{Q}}
  &= \xy{P}\left[\xpy{Q},\sgg \xy{Q},\xmy{Q}^2\right] - \frac{1}{2} \xmy{Q}
\end{aligned}\label{eq:componentwise_3d}
\end{equation}
with functions $ \xmy{P} $, $ \xpy{P} $, and $ \xy{P} $
analytic in their arguments,
where $Q_{\pm} := Q_{11} \pm Q_{22}$.
Note that the equations~\eq{eq:componentwise_3d}
are not analytic at $\sg=0$.
This is because $\sg$ is the only timescale,
so $\sg = 0$ corresponds to a singular case in which nothing evolves at all.
The same of course applies for more general flows in \eq{eq:Handeq_rate_indep}.

We see from \equ{eq:componentwise_3d}
that, when fitting these equations to simulation data, the basis of choice  is
$\{\xpy{Q},\xmy{Q},\xy{Q}\}$,
rather than the na\"{i}ve basis $\{Q_{11},Q_{22},\xy{Q}\}$.
Choosing the $\{\xpy{Q},\xmy{Q},\xy{Q}\}$ basis when plotting fits
has the effect of highlighting the important features of the data that
a given model is trying to fit.
For reversal protocols in particular,
apparent qualitative agreement between model and data for $Q_{11}(\gamma)$
and $Q_{22}(\gamma)$ can be exposed as clear disagreement in this basis.

%% file: linear_and_quadratic.tex

\section{Linear and quadratic Hand equations}\label{sec:results}

\subsection{Linear Hand equation: a no-go theorem}
We begin by assuming the fabric evolves via a frame-indifferent ODE linear in $\QQ$.
From \equ{eq:Handeq_general}, the most general linear model is
\begin{multline}
  \sgg \dgam{\QQ}
  =  \tsor{\hat{W}}\cdot  \QQ - \QQ\cdot  \tsor{\hat{W}}
  +\alpha_1 \QQ
  +\alpha_2 \tsor{\hat{E}}
  +\alpha_4 \tsor{\hat{E}}^2 \\
  +\alpha_5  \left( \tsor{\hat{E}} \cdot \QQ + \QQ \cdot \tsor{\hat{E}} \right)
  +\alpha_7 \left( \tsor{\hat{E}}^2 \cdot \QQ + \QQ \cdot \tsor{\hat{E}}^2 \right) \\
  - \frac{1}{3} \alpha_4 \Tr{\tsor{\hat{E}}^2}\Id
  -\frac{2}{3} \left( \alpha_5 \Tr{\QQ \cdot \tsor{\hat{E}}} + \alpha_7 \Tr{\QQ \cdot \tsor{\hat{E}}^2}\right) \Id
\end{multline}
with
\begin{align*}
  \alpha_1 &= x_{10} \\
  \alpha_2 &= x_{20} + x_{27} \hat{I}_7 + x_{29} \hat{I}_9 \\
  \alpha_4 &= x_{40} + x_{47} \hat{I}_7 + x_{49} \hat{I}_9 \\
  \alpha_5 &= x_{50} \\
  \alpha_7 &= x_{70}, \\
\end{align*}
in which we denote by $x_{ij}$ the constant coefficients
in $\alpha_i$ in front of $\hat{I}_j$
(with the exception of $x_{i0}$ which comes alone).
We recall that $\hat{I}_7 = \Tr{\QQ \tsor{\hat{E}}}$, and
$\hat{I}_9 = \Tr{\QQ \tsor{\hat{E}}^2}$.

As previously indicated, in simple shear flow, not all of the $x_{ij}$ can be separately tested against numerics.
In the non-redundant componentwise basis \eq{eq:componentwise_3d},
the linear model corresponds to fuctions $\xmy{P}$, $\xpy{P}$ and $\xy{P}$ of the form
\begin{align}\label{eq:linear_poly}
\xmy{P} &= \xmy{a} \\
\xpy{P} &= \xpy{a} + \xpy{b} \xpy{Q}
+ \xpy{c} \, \sgg \xy{Q} \\
\xy{P} &= \xy{a} + \xy{b} \xpy{Q} + \xy{c} \, \sgg \xy{Q},
\end{align}
where we have introduced seven independent constants (the $a$'s, $b$'s and $c$'s).

Interestingly this model is strongly constrained for the dynamics of $\xmy{Q}$.
More precisely, we now establish that equations~\eq{eq:componentwise_3d} and \eq{eq:linear_poly}
imply a rigorous condition on the post-reversal behavior of $\xmy{Q}$.
After reversal, $\QQ$ reaches a steady state value $\QQ^\pSS$
whose components, from \equ{eq:componentwise_3d} and \equ{eq:linear_poly}, satisfy
\begin{equation}
  0 = \xmy{a}\sgg \, \xmy{Q^\pSS} + 2 \xy{Q^\pSS}.
\end{equation}
Subtracting this from the first line of~\equ{eq:componentwise_3d} and
defining $\Updelta \QQ := \QQ - \QQ^\pSS$, we get
\begin{equation}
  \xmy{\Updelta \dgam{Q}} = \xmy{a}\sgg \, \xmy{\Updelta Q} + 2 \xy{\Updelta Q},
\end{equation}
Using the property of the pre-reversal steady state that
$\xmy{Q}(\gamma=0) = \xmy{Q}^\mSS = \xmy{Q}^\pSS$, this can be integrated as
\begin{equation}\label{eq:nogo_theorem}
\Updelta \xmy{Q}(\gamma) = 2 \int_0^{\gamma} e^{\xmy{a} \sgn(\gamma-\gamma')}  \Updelta\xy{Q}(\gamma') \mathrm{d}\gamma'.
\end{equation}
From this, we can see in particular that until $\Updelta \xy{Q}(\gamma)$ changes sign, the sign of
$\Updelta \xmy{Q}(\gamma)$ has to stay the same as $\sgn(\gamma) \sgn \left(\Updelta \xy{Q}(0) \right)$,
irrespective of the values of the $a$, $b$ and $c$ coefficients. This is a strong prediction for \emph{any} linear model
that can be easily compared with the numerical data.

\begin{figure}
  \centerline{\includegraphics[width=0.99\textwidth]{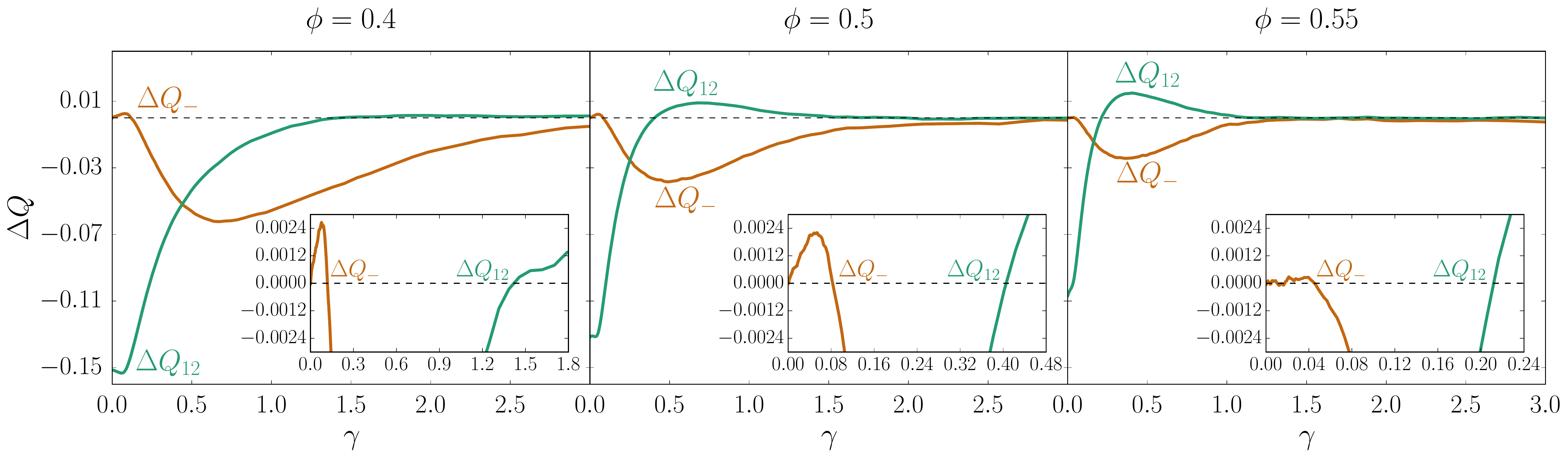}}
  \caption{Simulation data for $\Updelta \xmy{Q}$ and $\Updelta \xy{Q}$ against
post-reversal strain for three volume fractions $\phi=0.4$ (left), $\phi=0.5$ (center) and $\phi=0.55$ (right).
Any fabric evolution model which is linear in the fabric itself predicts that $\Updelta \xmy{Q}$ cannot change sign before $\Updelta \xy{Q}$ does,
from \equ{eq:nogo_theorem}.
This condition is unambiguously violated for all of our simulation data.
Insets: zoom in on the region where the condition is violated.
\label{nogofig}}
\end{figure}

Figure~\fig{nogofig} shows the post-reversal evolution of
$\Updelta \xmy{Q}$ and $\Updelta \xy{Q}$ for the three volume fractions we investigated.
In every case,
$\Updelta \xmy{Q}$ changes sign long before $\Updelta \xy{Q}$.
We moreover verified that this behavior is observed for any coarse-graining length $\epsilon$.
This allows us to conclude that \emph{no linear model can yield a good
fit to the data}.
This is the first important conclusion of our work.
We are not aware of any similar rejection of linear fabric evolution models
in previous work that does not address strain reversal.

Beyond ruling out the possibility of finding a suitable linear
model, the simulations allow us to conduct diagnostics,
identifying the source of the difficulty. $\QQ$
evolves between $\gamma$ and $\gamma+\mathrm{d}\gamma$ through three distinct
processes: the advection of pairs of particles with separation
$h<\epsilon$; the birth of near-contacts counted in
$\QQ$ when a pair of particles has separation
$h>\epsilon$ at strain $\gamma$ but $h<\epsilon$
an interval $\gamma+\mathrm{d}\gamma$ later;
and the death of near-contacts in the opposite case.
Writing $r^b$ and $r^d$
for respectively the rate of near-contact births and deaths per near-contact,
$\QQ^b$ and $\QQ^d$ for the fabric of near-contact births and deaths,
and $\dgam{\QQ}^a$ for the rate of change of $\QQ$ due to the
advection of near-contacts not instantaneously being
born or dying, we have (see Appendix~\ref{Q_decomp})
\begin{equation}
\dgam{\QQ} = \dgam{\QQ}^a + r^b \QQ^b - r^d \QQ^d. \label{Qbd}
\end{equation}

\begin{figure}
  \centerline{\includegraphics[width=0.99\textwidth]{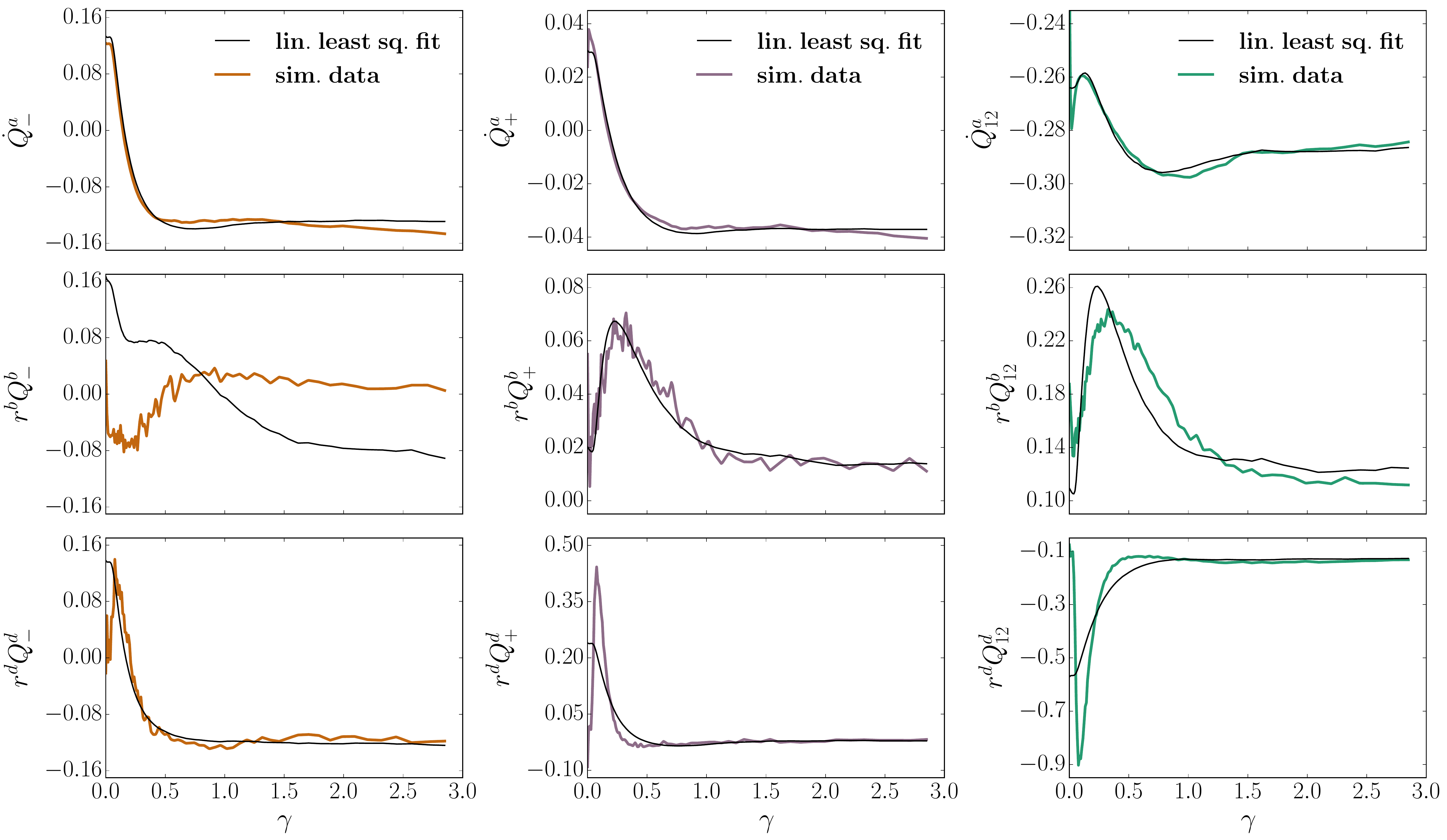}}
  \caption{The three $\dgam{\QQ}$ contributions $\dgam{\QQ}^a$ (top), $r^b \QQ^b$ (middle), and $r^d \QQ^d$ (bottom),
  as defined by~\equ{Qbd}, as a function of strain. For each contribution the three independent components
  $\xmy{\dgam{Q}^a}$, $\xpy{\dgam{Q}^a}$ and $\xy{\dgam{Q}^a}$
  (resp. $r^b\xmy{\QQ^b}$, $r^b\xpy{\QQ^b}$ and $r^b\xy{\QQ^b}$ and
  $r^d\xmy{\QQ^d}$, $r^d\xpy{\QQ^d}$ and $r^d\xy{\QQ^d}$) are shown resp. in the left, center and right columns (black lines).
  Each component is compared to a fit to the linear model \equ{eq:linear_poly} (colour lines).
\label{linadvfig}}
\end{figure}

Instead of considering directly $\dgam{\QQ}$ as a function of $\QQ$,
we can consider each term $\dgam{\QQ}^a$, $r^b \QQ^b$ and $r^d \QQ^d$ separately as a function of $\QQ$.
Trying to fit $\dgam{\QQ}^a$ as a function of $\QQ$ from simulation data  with the linear model in \equ{eq:componentwise_3d},
we obtain an excellent result, shown in the top row of \figu{linadvfig}.
This implies that the need for a non-linear
model is not due to the advective part of the evolution,
but rather due to the birth and death of near-contacts.
This is confirmed in the middle and bottom row of \figu{linadvfig}, showing that
the linear model fails for both the birth and death contributions, and gives
particularly bad results when the birth and advective contributions take their largest values
in amplitude, i.e. when a good accuracy matters most.

\subsection{Quadratic Hand equation}
Having ruled out the possibility of an adequate linear model, we
consider the general quadratic model
\begin{equation}
  \begin{aligned}
  \xmy{P} & =
    \xmy{a}
      + \xmy{b} \xpy{Q}
      + \xmy{c} \, \sgg \xy{Q}\\
  \xpy{P} & =
    \xpy{a}
      + \xpy{b} \xpy{Q}
      + \xpy{c} \, \sgg \xy{Q}
      + \xpy{d} \xpy{Q}^2 + \xpy{e} \xy{Q}^2
      + \xpy{f} \xmy{Q}^2 \\
  \xy{P} & =
    \xy{a}
      + \xy{b} \xpy{Q}
      + \xy{c} \, \sgg \xy{Q}
      + \xy{d} \xpy{Q}^2
      + \xy{e} \xy{Q}^2
      + \xy{f} \xmy{Q}^2.
\end{aligned}\label{eq:quadratic_model}
\end{equation}

The parameter space of this model, with 15 dimensions, is difficult to explore.
The strategy we adopt can be found in section 5.1 of \citet{Cheng2007},
and is to numerically differentiate $\QQ$
(see Appendix~\ref{Q_decomp}),
and use linear least squares to fit these quadratic models, with $\QQ$ taken
from simulation data, to $\dgam{\QQ}$.
The logic behind this is that,
if the model is capable of yielding a good qualitative fit to $\QQ$,
it will also provide an adequate fit to $\dgam{\QQ}$,
particularly if the fit is quantitatively good as well.

This approach very often fails for quadratic models
governed by \equ{eq:quadratic_model},
because the parameter set obtained in this way may,
when used to evolve $\QQ$ from its initial condition,
cause $\QQ$ to grow unbounded.
It is easy to see why: quadratic models correspond to the overdamped dynamics of
a three-dimensional vector
$\mathbf{s} = \{\xmy{Q}, \xpy{Q}, \xy{Q}\}$ in a cubic potential
$\dot{s}_i = -\nabla_{s_i}[A_{jkl}s_j s_k s_l + B_{jk}s_j s_k + C_{k}s_k]$,
with $\tsor{A}$, $\tsor{B}$ and $\tsor{C}$ tensors depending
on the coefficients of \equ{eq:quadratic_model}.
A cubic potential is generically not confining, so that
 unless the initial conditions are within the basin of attraction of a local potential minimum,
$\mathbf{s}$ will grow unbounded.
Finding a well-behaved quadratic fabric evolution model then involves identifying parameters
for which $\{\xmy{Q}^\mSS, \xpy{Q}^\mSS, \xy{Q}^\mSS\}$ lies within such a basin.
This appears as an unsatisfactory approach at best.

\begin{figure}
  \centerline{\includegraphics[width=0.99\textwidth]{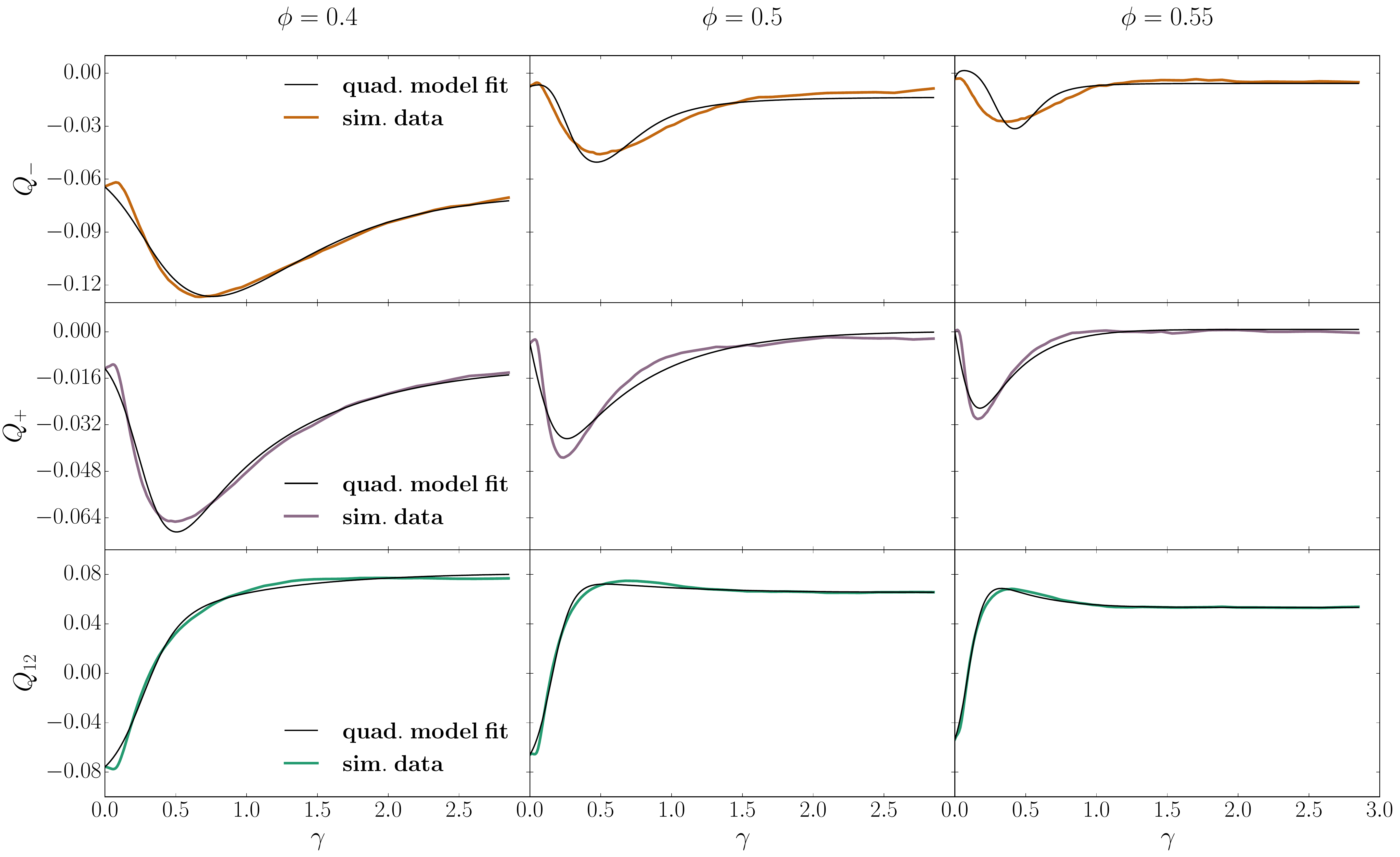}}
  \caption{Fit of the quadratic model
described by \equ{eq:subquad} to simulation data for the three components
of $\QQ$ (respectively $\xmy{Q}$, $\xpy{Q}$ and $\xy{Q}$ from top to bottom)
against post-reversal strain for $\phi=0.4$ (left), $\phi=0.5$ (center) and $\phi=0.55$ (right).
\label{quadfitsfig}}
\end{figure}

By subtracting some parameters from~\ref{eq:quadratic_model},
it is possible to restrict the parameter space enough
to find a model with a bounded evolution.
For instance, the best quadratic model we could find takes the form:
\begin{equation}
  \begin{aligned}
  \xmy{P}
  &= \xmy{a}
  + \xmy{b} \xpy{Q}
  + \xmy{c} \, \sgg \xy{Q} \\
  \xpy{P}
  &= \xpy{a}
  + \xpy{b} \xpy{Q}
  + \xpy{c} \, \sgg \xy{Q}
  + \xpy{e} \xy{Q}^2 \\
  \xy{P}
  &= \xy{a}
  + \xy{b} \xpy{Q}
  + \xy{c} \, \sgg \xy{Q}
  + \xy{e} \xy{Q}^2. \label{eq:subquad}
  \end{aligned}
\end{equation}
This contains 11 free parameters,
which we fit initially by linear least squares on \equ{eq:componentwise_3d},
that is, minimizing over the parameters
$\xmy{a}$, $\xmy{b}$, $\xmy{c}$, $\xpy{a}$, $\xpy{b}$, $\xpy{c}$, $\xpy{e}$,  $\xy{a}$, $\xy{b}$, $\xy{c}$ and $\xy{e}$
the quantity
\begin{equation}
    \begin{split}
        X_1 = \int_{\gamma=0}^{\gamma=3} \mathrm{d} \gamma & \left(\xmy{\dgam{Q}} - \xmy{P}\left[ \xpy{Q},\sgg \xy{Q}, \xmy{Q}^2 \right] \sgg \, \xmy{Q}
        - 2 \xy{Q}\right)^2 \\
        & + \left( \xpy{\dgam{Q}} - \xpy{P}\left[ \xpy{Q},\sgg \xy{Q}, \xmy{Q}^2 \right] \sgg \right)^2 \\
        & + \left(\xy{\dgam{Q}} - \xy{P}\left[\xpy{Q},\sgg \xy{Q},\xmy{Q}^2\right] + \frac{1}{2} \xmy{Q}\right)^2
    \end{split}\label{eq:initial_least_square}
\end{equation}
with strain-series $\QQ^\mathrm{data}$ and $\dgam{\QQ}^\mathrm{data}_{ij}$ taken from the simulation data.
Note that there is some freedom in choosing the upper $\gamma$ limit in the integral
of \equ{eq:initial_least_square}; a larger maximum $\gamma$ will favour fits that are accurate in steady state,
while a smaller one will favour a good accuracy in the early phase.
The value $\gamma=3$ picked here is the one which we found to give the most balanced results between both extremes.
After this initial fit, we further apply a gradient descent to minimise the linear least squares
between the actual model prediction $\QQ^\mathrm{fit}$
(obtained by integrating \equ{eq:componentwise_3d} over strain with
initial conditions from the simulation data $\QQ^\mathrm{data}(\gamma=0)$)
and the numerical data $\QQ^\mathrm{data}$, that is, minimizing
\begin{equation}
    \begin{split}
        X_2 = \int_{\gamma=0}^{\gamma=3} \mathrm{d} \gamma
                               \left(\xmy{Q^\mathrm{fit}} - \xmy{Q^\mathrm{data}} \right)^2
                              + \left(\xpy{Q^\mathrm{fit}} - \xpy{Q^\mathrm{data}} \right)^2
                               + \left(\xy{Q^\mathrm{fit}} - \xy{Q^\mathrm{data}} \right)^2
    \end{split}\label{eq:final_least_square}
\end{equation}
to get final best-fit parameters.

\begin{figure}
  \centerline{\includegraphics[width=0.99\textwidth]{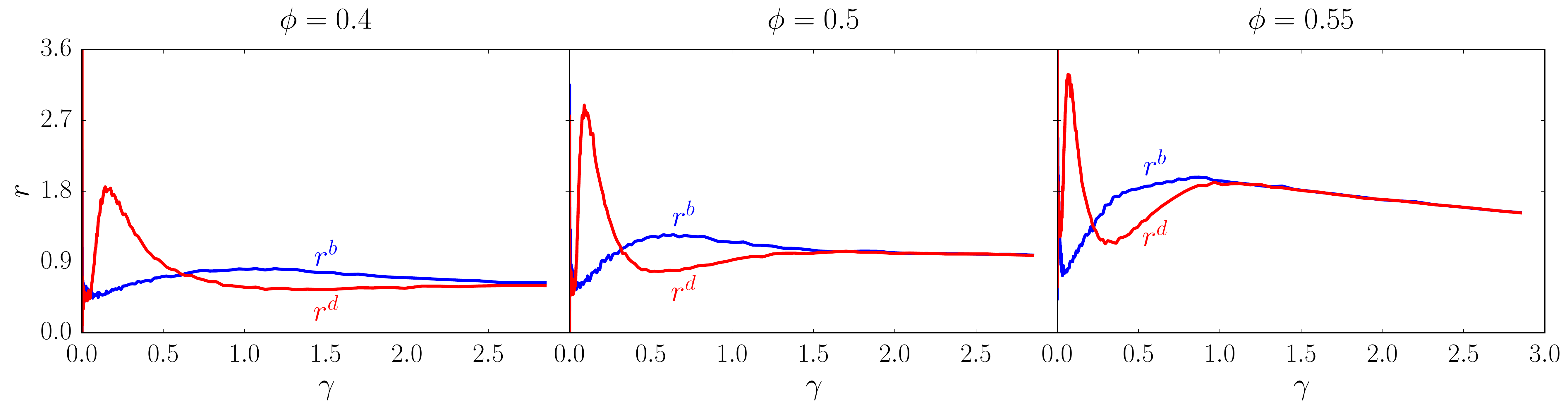}}
  \caption{Fractional birth rate $r^b$ (blue) and death rate $r^d$ (red)
against post-reversal strain for $\phi=0.4$ (left), $\phi=0.5$ (center) and $\phi=0.55$ (right).
\label{bdratesfig}}
\end{figure}

Depending on the component of interest,
this procedure leads to fits ranging from good to excellent, as shown in \figu{quadfitsfig}.
The only obvious weakness is a moderate but increasing discrepancy for $\xmy{Q}$ when $\phi$ increases.
This is likely due to the smaller $r^b$ and $r^d$ at low volume fraction
(see \figu{bdratesfig}), yielding flatter, easier-to-fit curves.

Because of the difficulty to find a well-behaved model,
we cannot perform a thorough exploration of parameter space,
and there may be a better fitting quadratic model than \equ{eq:subquad}.
In any case the fitting power of quadratic models is, in three dimensions, in principle satisfactory when it comes to matching the numerically observed fabric evolution. However we believe this to be largely coincidental, rather than faithful to the underlying physics, and stemming simply from the large number of parameters available. This is partly because there is no reason to expect the parameters of a mechanistically faithful model to need fine tuning to avoid the generic blow-up described previously. However, a stronger reason to reject such models is found by making a detour to the two-dimensional case, as we describe next.

%% file: 2d_insight.tex

\section{Insights from the two-dimensional case}\label{sec:2d_insight}

As we have shown in section~\ref{sec:results}, linear models
do not contain enough physics to capture the relation
between the post-reversal dynamics of $\xmy{Q}$ and $\xy{Q}$,
that is, the dynamics in the shear plane.
Quadratic models are able to achieve quantative fit to $\xmy{Q}$ and $\xy{Q}$,
but at the price of a greatly increased number of free parameters which, coupled with the generic presence of instabilities,
suggests that the difficulty has been circumvented
simply by over-parameterizing the problem.
To understand better what is the source of the difficulty,
we can consider the simplified
case of a two dimensional suspension,
for which by construction the shear plane dynamics can be
described in isolation, with the hope that it shares
its characteristic features with the three-dimensional case.

\begin{figure}
  \centering
  \includegraphics[width=0.99\textwidth]{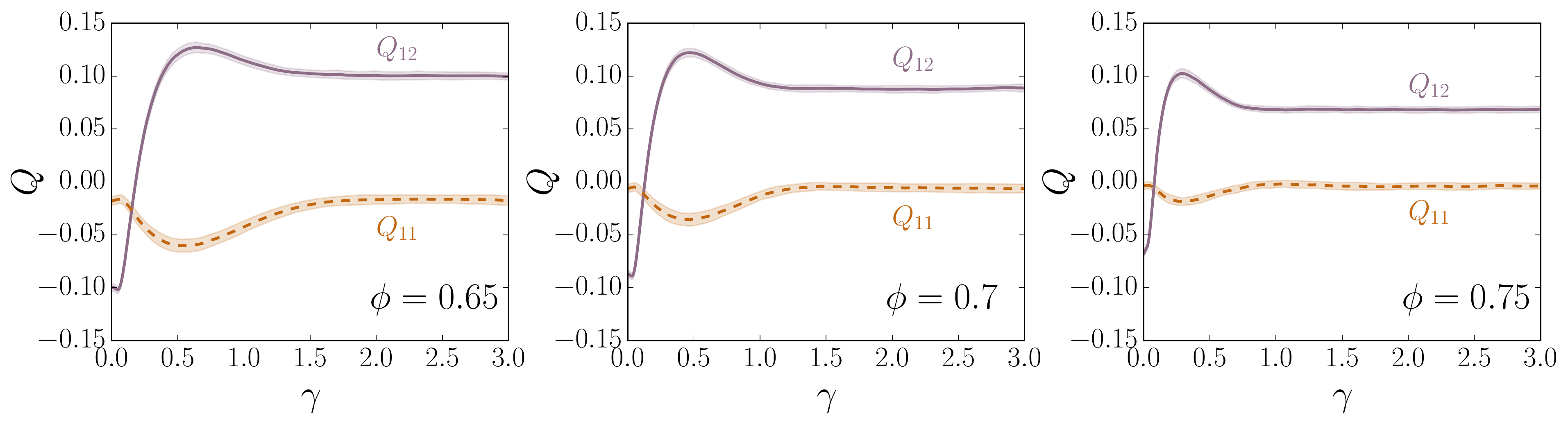}
  \caption{Fabric tensor components upon reversal at $\gamma=0$ for $\phi=0.65, 0.7$ and $0.75$
  in two dimensions. As in three dimensions, $\QQ$ flips on a strain scale of order 1,
  there is an overshoot in $\xy{Q}$ and $\xmy{Q}$ ($=2Q_{11}$ in two dimensions) is negative.
  Also, $\QQ$ gets smaller and flips on a shorter strain scale as $\phi$ increases.
  Thick dark-shaded lines are the averaged data, while the light shaded area around each curve
  is the standard deviation obtained from the individual shear reversals.
  Note that the variance of the individual runs appears smaller here than
  in the three-dimensional case in \figu{fig:Q_numerics_delta_dep}.
  This is the result of using larger system sizes in two dimensions ($N=4000$) than in three ($N=500$).}\label{fig:2d_reversal}
\end{figure}

Thus, in this section we perform numerical simulations of a bidimensional monolayer of spheres,
and we try to model the evolution of the fabric tensor, as before.
The simulation technique is the same as in three dimensions, only with a different
number of particles $N=4000$ and different area fractions $\phi=0.65, 0.7$ and $0.75$.
Also, the average results for the fabric tensor are obtained over $100$ shear reversal realizations.

In two dimensions, the fabric tensor $\QQ := \langle \vec{p}\vec{p} \rangle -(1/2)\tsor{I}$
reduces to two independent non-zero components $\xy{Q}$ and $\xmy{Q}=2Q_{11}=-2Q_{22}$.
The fabric evolution in two dimensions is shown in \figu{fig:2d_reversal} for $\phi=0.65$, $0.7$ and $0.75$.
Comparing these data with the three-dimensional ones shown in \figu{fig:Q_delta_phi_dep},
we can see that the qualitiative behavior is the same in two and three dimensions,
and we can thus gain insight from the simplified two-dimensional case.

\begin{figure}
  \centerline{\includegraphics[width=0.99\textwidth]{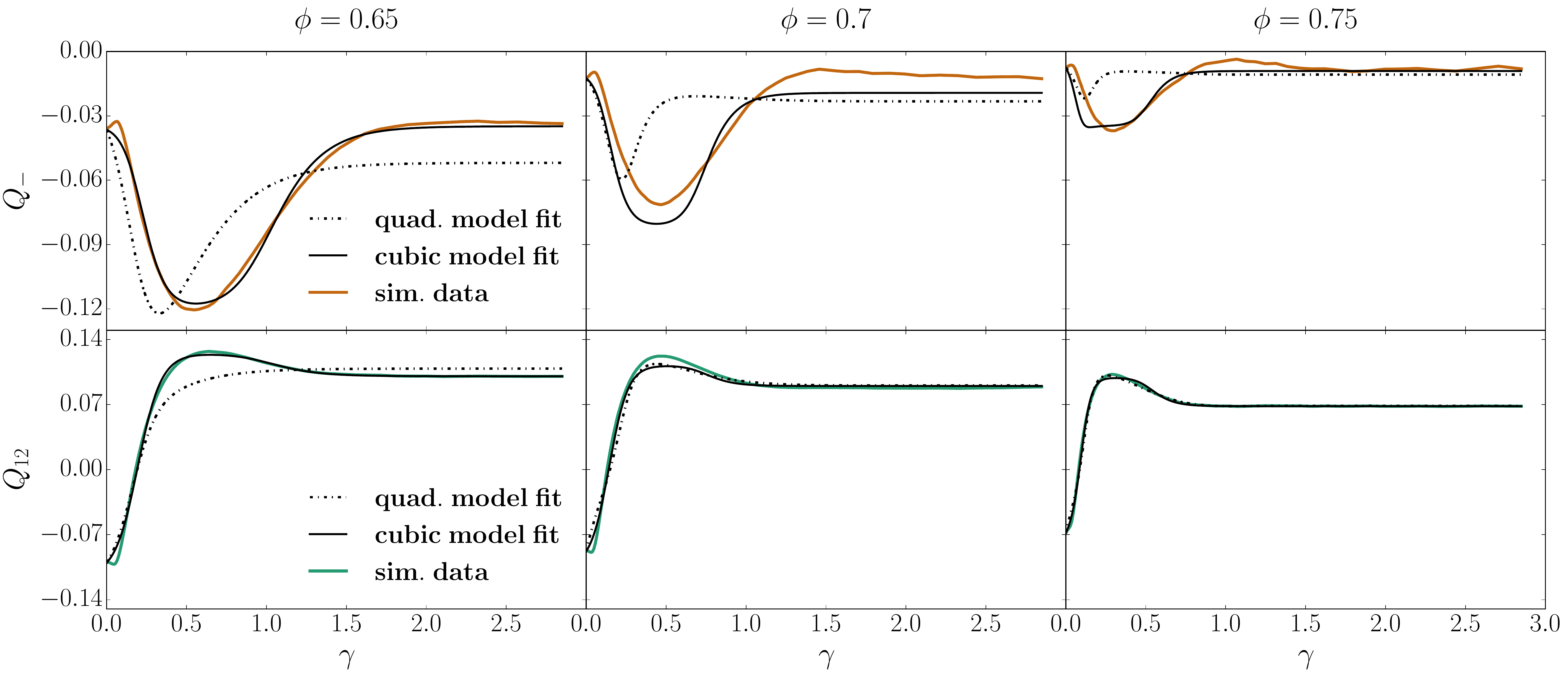}}
  \caption{Fits of the quadratic model described by \equ{eq:2d_quad}
and the cubic model described by \equ{eq:2d_subcube}
to the two-dimensional fabrics data against post-reversal strain,
for both components $\xmy{Q}$ (top) and $\xy{Q}$ (bottom), and for three area fractions
$\phi=0.65$ (left), $\phi=0.7$ (center) and $\phi=0.75$ (right).
\label{fits2dfig}}
\end{figure}

The two-dimensional version of the system of ordinary
differential equations~\eq{eq:componentwise_3d} is
(see Appendix~\ref{Handapp})
\begin{equation}
  \begin{aligned}\label{eq:componentwise_2d}
    \xmy{\dgam{Q}} &= \xmy{P}\left[ \sgg \xy{Q}, \xmy{Q}^2 \right] \sgg \xmy{Q} + 2 \xy{Q}, \\
    \xy{\dgam{Q}} &= \xy{P}\left[ \sgg \xy{Q}, \xmy{Q}^2 \right]
    - \frac{1}{2} \xmy{Q}
\end{aligned}
\end{equation}

As in three dimensions, we are able to conclude that a linear
model is inadequate: the same no-go theorem holds (recall Section~\ref{sec:results}),
and the data unambiguously violate it.
Turning now to the quadratic model, this is considerably simpler than in three dimensions.
Where there were 15 free parameters in three dimensions, the two-dimensional model contains only 6, in terms of which
\begin{equation}
  \begin{aligned}
  \xmy{P}
  &= \xmy{a}
  + \xmy{c} \, \sgg \xy{Q}\\
  \xy{P}
  &= \xy{a}
  + \xy{c} \, \sgg \xy{Q}
  + \xy{e} \xy{Q}^2
  + \xy{f} \xmy{Q}^2.\label{eq:2d_quad}
\end{aligned}
\end{equation}

In \figu{fits2dfig}, we show that while leading to qualitatively correct fits,
the quadratic model does not provide a quantitative fit.
In particular, the minimum in $\xmy{Q}$ is quite poorly captured,
both in position and amplitude, and this discrepancy is accentuated at higher $\phi$.
Because of the qualitative similarity between the data in two and three dimensions,
we can expect that the three-dimensional fabric dynamics in the shear plane
is essentially given by the two-dimensional one, perhaps augmented by a weak coupling
to the vorticity direction through a $\xpy{Q}$ term.
Indeed, it is hard to imagine a mechanistic interpretation of any quadratic model
that would not be expected to work just as well in two dimensions.
The fact that \equ{eq:2d_quad} does not describe even semi-quantitatively the dynamics in two dimensions
is telling us that the quadratic model in \equ{eq:subquad}
owes its moderate quantitative success to the mathematical freedom of having 11 free parameters
rather than to its capturing the underlying physics.

To achieve a near-quantitative fit for the two-dimensional data,
we must instead go to cubic order with the model
\begin{equation}
  \begin{aligned}
  \xmy{P}
  &= \xmy{a}
  + \xmy{c} \, \sgg \xy{Q}
  + \xmy{e} \xy{Q}^2
  + \xmy{f} \xmy{Q}^2\\
  \xy{P}
  &= \xy{a}
  + \xy{c} \, \sgg \xy{Q}
  + \xy{e} \xy{Q}^2
  + \xy{f} \xmy{Q}^2. \label{eq:2d_subcube}
\end{aligned}
\end{equation}
(This is a cubic model in $\QQ$ because $\xmy{P}$ multiplies $\xmy{Q}$ in~\equ{eq:2d_quad}.)
The data-fitting approach described in the previous section,
of using linear least squares on
numerically-differentiated $\QQ$ data to obtain an initial guess
for gradient descent, does not work for this cubic model.
Indeed, the problem encountered for the three-dimensional quadratic model in the previous
section gets even more severe for \equ{eq:2d_subcube},
that is, the vast majority of parameter sets entering \equ{eq:2d_subcube}
will lead to an unbounded time evolution for $\QQ$, because of the non-confining character
of the resulting ODE.
This is a commonplace feature of high order polynomial models with insufficiently constrained parameters.
We instead obtain quantitatively superior fit by
initiating a gradient descent for cubic model from the best fit
parameters for the quadratic model of
\equ{eq:2d_quad}.
This cubic model can fit the data with a reasonable accuracy, as shown in \figu{fits2dfig},
although the fits to the $\xmy{Q}$ data show room for improvement.
More worryingly, signs of overfitting are clearly visible in the $\xmy{Q}$ component.
In particular some short strain scales features close to the minimum at intermediate
strain values around $\gamma\approx 0.5$ appear in the fit.

\begin{figure}
  \centerline{\includegraphics[width=0.99\textwidth]{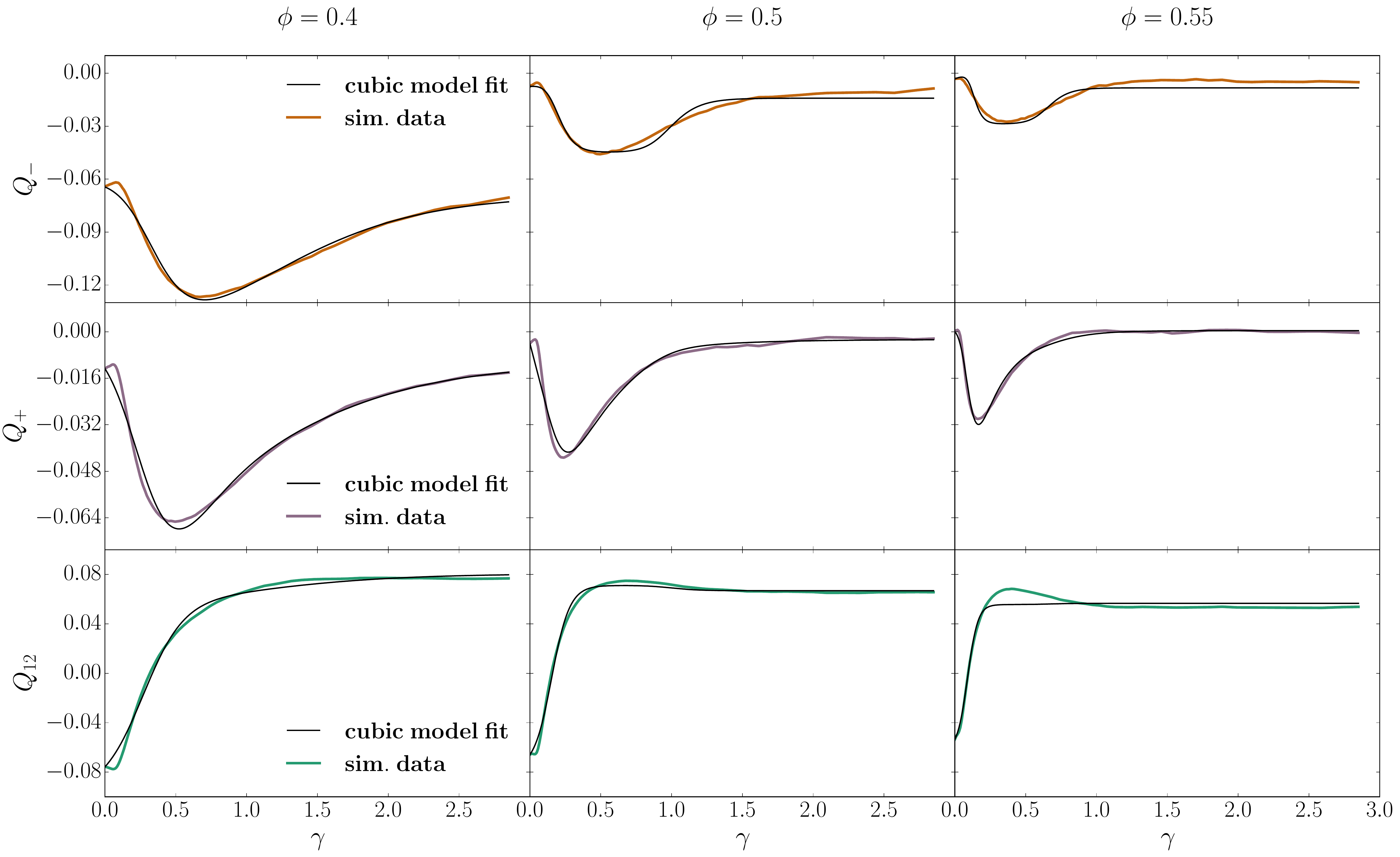}}
  \caption{Fit of the cubic model described by \equ{eq:3d_subcube} to three-dimensional simulation data
  for the components $\xmy{Q}$ (top), $\xpy{Q}$ (middle), $\xy{Q}$ (bottom) against post-reversal strain
  for $\phi=0.4$ (left), $\phi=0.5$ (center) and $\phi=0.55$ (right).
\label{cubefitsfig}}
\end{figure}

Empirically, we can create a three-dimensional cubic model
\begin{equation}
  \begin{aligned}
  \xmy{P}
  &= \xmy{a}
  + \xmy{c} \, \sgg \xy{Q}
  + \xmy{e} \xy{Q}^2
  + \xmy{f} \xmy{Q}^2  \\
  \xpy{P}
  &= \xpy{a}
  + \xpy{b} \xpy{Q}
  + \xpy{c} \, \sgg \xy{Q}
  + \xpy{e} \xy{Q}^2 \\
  \xy{P}
  &= \xy{a}
  + \xy{c} \, \sgg \xy{Q}
  + \xy{e} \xy{Q}^2
  + \xy{f} \xmy{Q}^2 \label{eq:3d_subcube}
\end{aligned}
\end{equation}
by supplementing the closed shear-plane dynamics of \equ{eq:2d_subcube}
with the vorticity dynamics of \equ{eq:subquad}.
Although this works quite well (see \figu{cubefitsfig}),
we feel it is unlikely that any mechanical insight
can be inferred from this exercise. This is in part due to the
large number of parameters involved, and in part due to the
smallness of the region in parameter space in which
$\QQ$ does not blow up when solving the model.

%% file: limitations.tex

\section{Fabric evolution models: limitations}\label{sec:discussion}
At this stage, we can conclude that any continuum model for
$\QQ$ in dense non-Brownian suspensions must be of a
different and more complicated form than many models for
non-Newtownian fluids~\citep{larson2013constitutive}.
Specifically: models linear in $\QQ$
are ruled out; models quadratic in $\QQ$ fail
in two dimensions and are therefore suspect in three;
and cubic models generically overfit the data
without mechanical insight.

To understand why this approach to finding closed evolution equations
for $\QQ$ breaks down in dense suspensions, we must revisit our basic assumptions, which are two-fold.
Firstly, we assumed in effect that at any instant $\QQ$ is an adequate representation of the full probability density
of near-contacts $P(\vec{p})$, that is,
$P(\vec{p})$ is well approximated by its second-order spherical harmonic expansion \eq{she2}.
In a second and closely related assumption, we postulated that $\QQ$ contains enough microstructural
information to obtain $\dgam{\QQ}$ for any given instantaneous strain rate $\tsor{K}$, without information
from higher moments of $P(\vec{p})$ or even more generally from the entire pair correlation function $g(\vec{r})$
(which includes radial as well as orientational information).
Put differently, we can find a closure in $\QQ$ that approximates how the higher moments of $\dot P(\vec{p})$
(higher spherical harmonics) contribute to $\dgam{\QQ}$.
In principle the second assumption does not require the first one, but it seems unlikely \emph{a priori}
that $\QQ$ determines higher-order spherical harmonic contributions to $P(\vec{p})$
when these are larger than the contribution of $\QQ$ itself.

\begin{figure}
  \centerline{\includegraphics[width=0.99\textwidth]{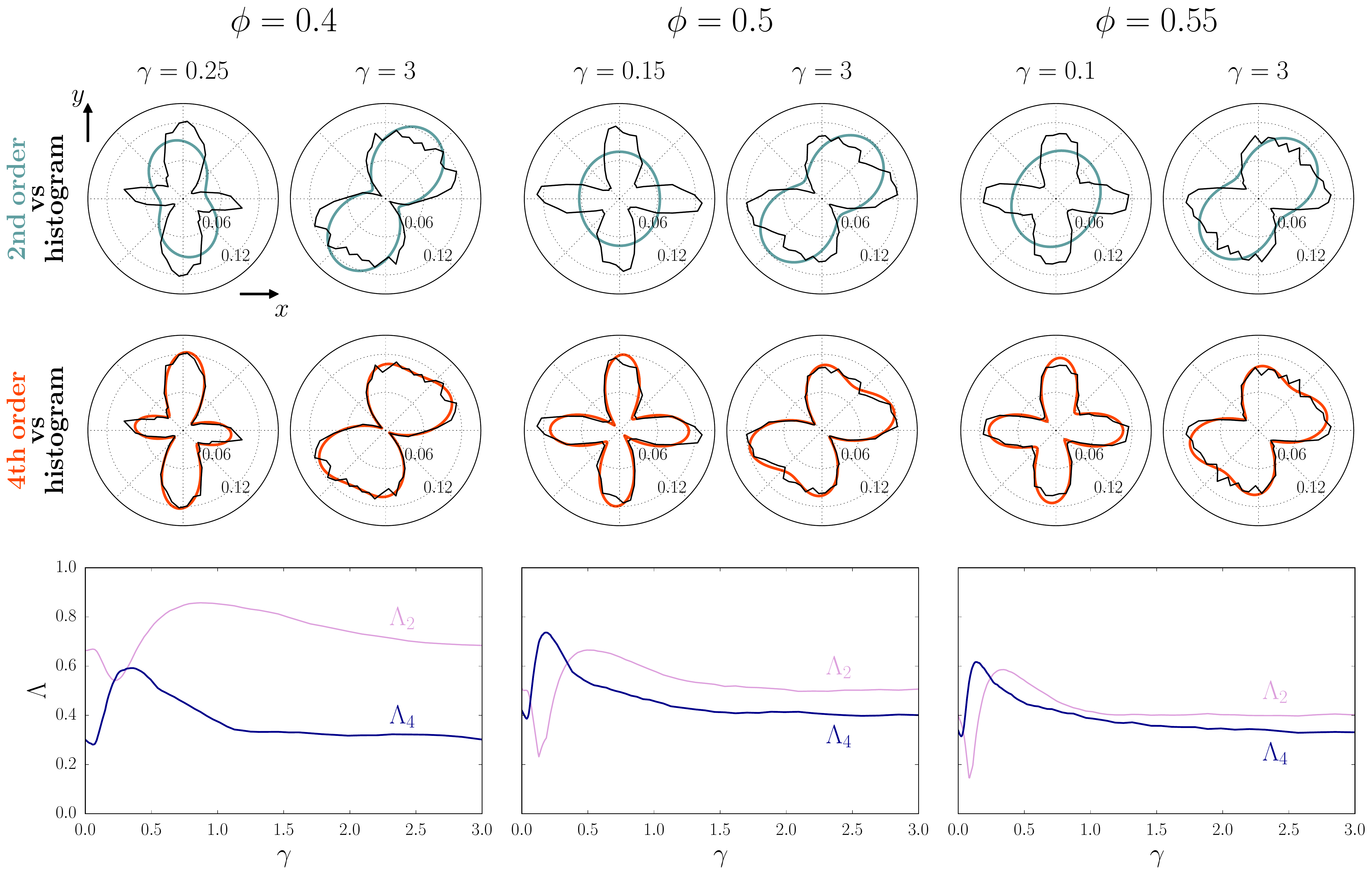}}
  \caption{Quantification of the ability of the fabric tensor $\QQ$
           to capture the major features of the full near-contact orientation distribution function $P(\vec{p})$
           during shear reversal for three volume fractions $\phi=0.4$ (left column),
           $\phi=0.5$ (middle column), and $\phi=0.55$ (right column).
           \textbf{Top row:} Polar plots of $P(\vec{p})$ in the shear plane (black line) compared to its second-order
           spherical harmonic approximation (\equ{she2}, blue line),
           for the two strain values after reversal indicated on top of each column for each $\phi$.
           \textbf{Second row:} Polar plots of the same $P(\vec{p})$ data, compared to its fourth-order
           spherical harmonic approximation (\equ{she4}, yellow line).
           \textbf{Bottom row:} Largest eigenvalues $\Lambda_2$ and $\Lambda_4$ of resp. the second-rank $\QQ$ and
           the fourth-rank $\tsor{C}$ (see main text for the definition) associated respectively to second-order and fourth-order contributions
           in the spherical harmonic expansion of $P(\vec{p})$.
\label{lobalityfig}}
\end{figure}

We show in this section that both these assumptions fail for at least parts of the shear reversal.
We first assess the adequacy of $\QQ$ as a description of the suspension microstructure.
In order to compare $P(\vec{p})$ to its successive low-order approximations in terms of spherical harmonics -- among which only those of even order contribute --
let us recall that the second-order approximation is given by \equ{she2} and the fourth-order one is~\citep{ken-ichi_distribution_1984}
\begin{equation}
P(\vec{p})\approx
\frac{1}{4 \pi} \left( 1 + \frac{15}{2} \QQ : \vec{p}  \vec{p}
+\frac{315}{8} \tsor{C} :: \vec{p}  \vec{p}  \vec{p}  \vec{p} \right), \label{she4}
\end{equation}
with a fourth-rank structure tensor $ \tsor{C} $ with components
\begin{equation}
C_{ijkl} = \langle p_i p_j  p_k p_l \rangle
- \frac{1}{7} H_{ijkl}
+ \frac{1}{35} I_{ijkl},
\end{equation}
where
\begin{align}
H_{ijkl} &= \langle p_i p_j \rangle \delta_{kl}  + \langle p_i p_k \rangle \delta_{jl} + \langle p_i p_l \rangle \delta_{jk}
+ \delta_{ij} \langle p_k p_l \rangle + \delta_{ik} \langle p_j p_l \rangle + \delta_{il} \langle p_j p_k \rangle \\
I_{ijkl} &= \delta_{ij} \delta_{kl} + \delta_{ik} \delta_{jl} + \delta_{il} \delta_{jk}.
\end{align}

In the top part of \figu{lobalityfig}, we evaluate the relative contributions of second
and fourth-rank spherical harmonics to $P(\vec{p})$ for three volume fractions and
two representative strains $\gamma$ after shear reversal.
The first strain, $\gamma=0.25, 0.15$ and $0.1$ respectively for $\phi=0.4, 0.5$ and $0.55$, is shortly after reversal
and close to the strain at which the principal axes
of the microstructure flip over to become the post-reversal compressional and extensional axes.
The second strain, $\gamma=3$, corresponds to a microstructure that has reached its post-reversal steady state.
In the latter case $P(\vec{p})$ is effectively two-lobed for all three volume fractions,
that is, it has two local maxima and two local mimima in the shear-plane.
In this case it is well approximated by its second-order spherical harmonic.
However, at smaller strains, during the reconstruction of the contact network after reversal,
$P(\vec{p})$ has a distinctive four-lobed structure (that is, four local maxima),
which can only be captured by the fourth-order spherical harmonic.
For all three volume fractions, at short strains (left column for each volume fraction in \figu{lobalityfig})
the second-order approximation is failing to capture the amplitude of the anisotropic features of $P(\vec{p})$,
but even worse it does not even pick the major lobes' direction.

We can quantify further the four-lobed nature of $P(\vec{p})$ by comparing the largest eigenvalues
$\Lambda_2$ and $\Lambda_4$ of $\frac{15}{2}\QQ$ and $\frac{315}{8} \tsor{C}$ respectively
(see \citet{Qi2006} for a reference on the eigenvalues
of fully symmetric high-order tensors).
The time evolution of these eigenvalues under shear reversal is shown in the bottom part of \figu{lobalityfig}
for the same three volume fractions.
Here again this reveals at short post-reversal strains an interval across which
$\tsor{C}$ plays a bigger role than $\QQ$.
Even outside this strain interval, $\Lambda_4$ and $\Lambda_2$
have the same order of magnitude, including for large strains back to steady-state.
Importantly though, the breakdown of \equ{she2}
is much more severe in transient flows such as reversal
than in steady state.

Nonetheless, an approximation of $P(\vec{p})$ as a function of $\QQ$ could be in order provided
that the contribution of the fourth-order spherical harmonics could be inferred
from the knowledge of the second-order one, that is,
provided that there is an accurate closure of $\tsor{C}$ in terms of $\QQ$.
That the fourth-order spherical harmonics contribute to $P(\vec{p})$
as much if not more than the second-order one makes this possibility unlikely though.
The only way to build a fourth-rank tensor out of $\QQ$ is by direct product of powers of $\QQ$,
hence a closure of $\tsor{C}$ must be a weighted sum of terms of the form  $\QQ^n\QQ^m$.
Knowing that $\QQ$ can be expressed as the sum of dyadic products
of its mutually orthogonal eigenvectors $\QQ = \sum_i \lambda_i \vec{e}_i\vec{e}_i$,
the contribution to $P(\vec{p})$ of a term $\QQ^n\QQ^m$ in $\tsor{C}$ is proportional to (through~\equ{she4})
$(\QQ^n:\vec{p}\vec{p})(\QQ^m:\vec{p}\vec{p})$.
In two dimensions (the argument readily extends to the shear plane in three dimensions),
if we call $\theta$ the angle between $\vec{p}$ and $\vec{e}_1$, then this contribution is
$\propto (\QQ^n:\vec{p}\vec{p}) (\QQ^m:\vec{p}\vec{p}) = \lambda_1^{m+n} \cos^4\theta + \lambda_2^{m+n} \sin^4\theta + (\lambda_1^m\lambda_2^n+\lambda_1^n\lambda_2^m) \cos^2\theta \sin^2\theta$.
This term is thus in general a four-lobed contribution to $P(\vec{p})$, but an overly constrained one:
the directions of the lobes are the principal axes of $\QQ$, i.e. $\vec{e}_1$ and $\vec{e}_2$.
Therefore, any closure of $\tsor{C}$ in $\QQ$ gives a $P(\vec{p})$ with maxima and minima along the directions of the eigenvectors of $\QQ$.
Looking back at $P(\vec{p})$ alongside $\QQ$ in the top row of \figu{lobalityfig} reveals that
at early strains the lobes of $P(\vec{p})$ are not aligned with the principal axes of $\QQ$.

This is a clear indicator that a constitutive model based on a closed ODE
for fabric evolution under flow cannot capture the physics of
near-contact network reconstruction following flow reversal.
The strong intrinsic fourth-order component of $P(\vec{p})$ is one of the key outcomes
of our analysis.
It deems as unphysical closed fabric evolution models
even if they may apparently fit simulation data well,
as such a model would base the time evolution on the second-order spherical harmonics
which is subdominant and having incorrect principal axes during a significant part of the time evolution after reversal.

\begin{figure}
  \centerline{\includegraphics[width=0.75\textwidth]{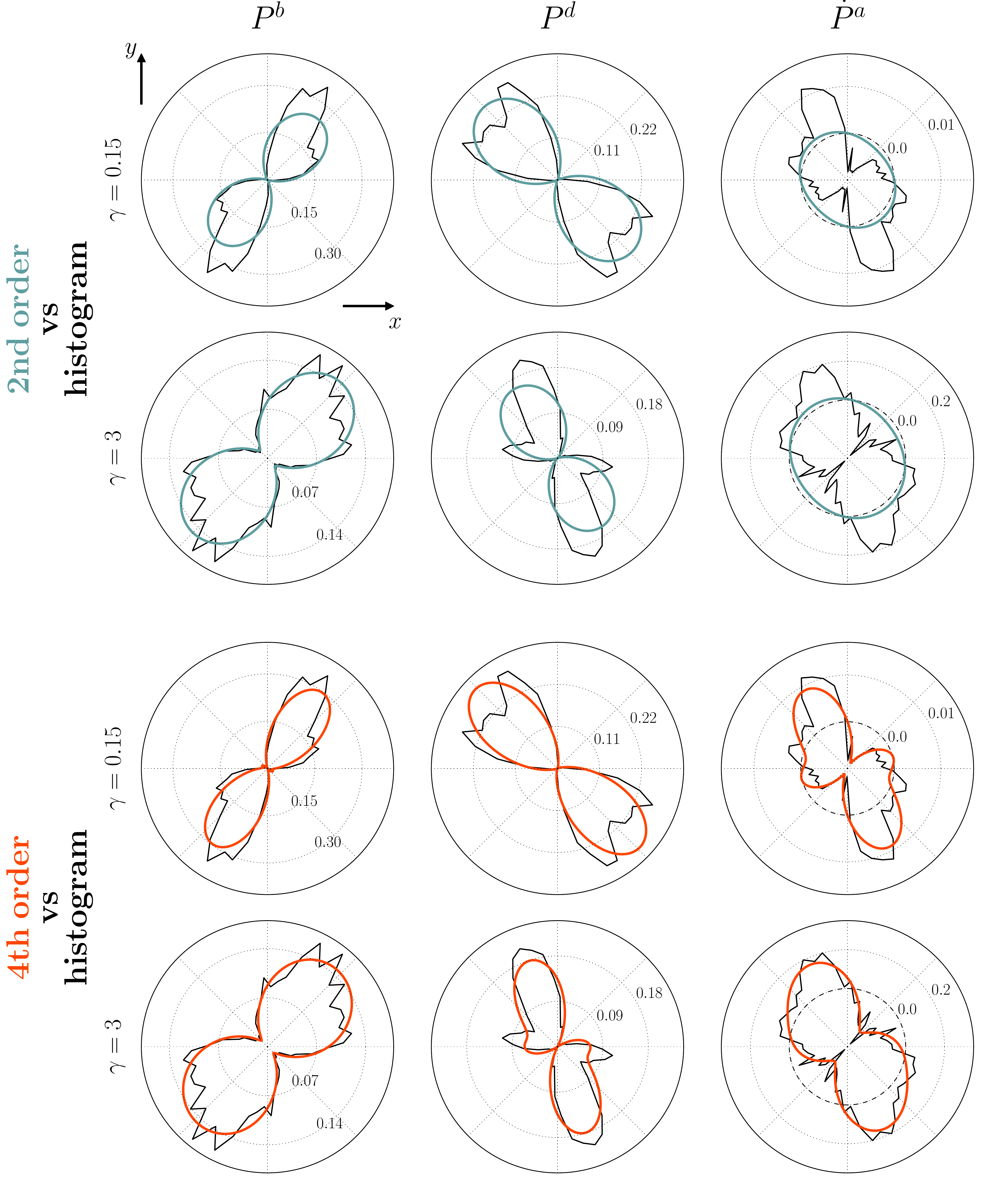}}
  \caption{\textbf{Top two rows:} Polar plots of $\dgam{P}^a(\vec{p})$ (left column),
  $P^b(\vec{p})$ (middle column) and $P^d(\vec{p})$ (right column),
  as defined in \equ{eq:Pdot_decomposition}, in the shear plane for $\phi=0.5$
  and for $\gamma=0.15$ (top row) and $3$ (second row),  in black lines.
  These are compared to their second-order spherical harmonic approximations (color lines).
  \textbf{Bottom two rows:} Same $\dgam{P}^a(\vec{p})$, $P^b(\vec{p})$ and $P^d(\vec{p})$ data (black lines),
  compared to their and fourth-order spherical harmonic approximations (color lines).
  \label{fig:Pdot_components}}
\end{figure}

A closer look at the strain derivative $\dgam{P}(\vec{p})$ confirms this view.
As for $\QQ$, we can decompose $\dgam{P}$ into advective, birth and death contributions:
\begin{equation}\label{eq:Pdot_decomposition}
\dgam{P}(\vec{p})
=\dgam{P}^a(\vec{p})
+r^b P^b(\vec{p})
-r^b P^d(\vec{p}),
\end{equation}
which is derived in Appendix~\ref{Q_decomp}
by considering the evolution of the number of near-contacts in a surface element on the
unit sphere.
Whereas $P^b(\vec{p})$ and $P^d(\vec{p})$ are positive quantities corresponding
to respectively the appearance and disappearance of near-contacts,
$\dgam{P}^a(\vec{p})$ will in general take positive and negative values in different directions.
The three contributions are shown in \figu{fig:Pdot_components} in the shear plane,
for $\phi=0.5$ and strains $\gamma=0.15$ and $\gamma=3$ (as in \figu{lobalityfig}),
alongside their second- and fourth-order spherical harmonics approximations.
(The data for $\phi=0.4$ and $0.5$ show similar features,
and are plotted in Appendix~\ref{sec:appendix_Pdot_comp_plot}.)

Surprisingly, except for $P^d(\vec{p})$ at steady state,
none of these distributions show marked four-lobed structures.
This is not an inconsistency with the fact
that $P(\vec{p})$ is four-lobed during part of the reversal though,
as it suffices that the contributions to $\dgam{P}(\vec{p})$ contribute
to accumulate (or deplete) near-contacts in different directions
to get a $P(\vec{p})$ with more than two maxima in the shear plane.

Nonetheless, as shown in the top rows of~\figu{fig:Pdot_components},
second-order spherical harmonics approximation fails to capture
some essential features of $P^b(\vec{p})$, $P^d(\vec{p})$ and $\dgam{P}^a(\vec{p})$.
Moreover, the worst failures occur when these components have large amplitudes relative to the others,
that is, at small strains for $P^b(\vec{p})$ and at large strains for $P^d(\vec{p})$ and $\dgam{P}^a(\vec{p})$.
Some features make these distributions particularly difficult to capture by a fabric-based approximation.
For instance, when $P^d(\vec{p})$ shows four lobes, the lobes
are consistently found not to be perpendicular to each other.
Similarly, even if $\dgam{P}^a(\vec{p})$ is bilobed, the directions associated with minima
are not orthogonal to the directions of the maxima, especially at smaller strains.
As discussed earlier in this section, these features cannot be captured by analytic functions of $\QQ$.
On the other hand, we show in the bottom of \figu{fig:Pdot_components} that
for all parameters $\dgam{P}^a(\vec{p})$, $P^b(\vec{p})$ and $P^d(\vec{p})$
are reasonably well approximated by their fourth-order spherical harmonic expansions.

Our detailed data for $P(\vec{p})$, as well as its time derivative, are thus unambiguous:
the angular distribution of near-contacts possesses a shear-plane structure only captured
by an expansion up to fourth-order in spherical harmonics.
As a consequence, the fabric evolution cannot be expressed only in $\QQ$ itself,
and contains information from the fourth-rank $\tsor{C}$ in a way that cannot be approximated
by an adequate closure.

%% file: discussion.tex

\section{Discussion}

Our detailed interrogation of the numerical data clearly exposes the challenge facing fabric evolution models,
whether considered in themselves or as the basis for a constitutive model for stress evolution under flow.
Their failure appears fundamental: the choice of a second-rank fabric tensor
to encode the essential features of the microstructural anisotropy is found wanting in shear reversal flows.
This is because the orientational distribution of the near-contact interactions (responsible for stress generation)
$P(\vec{p})$, and also its time derivative, strongly depart from the ellipsoidal structures
that are the only ones directly described by a second-rank tensor.

We emphasize that this is not a problem of choosing the \emph{wrong} second-rank tensor
to encode the microstructure.
Indeed, whereas the small anisotropy of $P(\vec{p})$ at high densities
motivates a low order spherical harmonic expansion in which $\QQ$ is the first nontrivial term,
other ans\"atze for $P(\vec{p})$ based on a symmetric second-rank tensor $\tsor{T}$ are possible. These include the Bingham distribution ansatz
$P(\vec{p}) \propto \exp(\tsor{T}:\vec{p}\vec{p})$~\citep{bingham_antipodally_1974,chaubal_closure_1998}.
In contrast to our \equ{she2}, these ans\"atze usually contain terms of all orders in the spherical harmonic expansion.
Nonetheless any $P(\vec{p}) \propto f(\tsor{T}:\vec{p}\vec{p})$, with $f$ a monotonic function,
is still a two-lobed distribution, and in consequence will not adequately describe four-lobed structures in the shear-plane,
let alone ones in which two pairs of lobes are oriented with non-perpendicular axes.
These characteristic four-lobed structures observed for $P(\vec{p})$ and $\dot P(\vec{p})$ in the shear-plane during significant parts
of the shear reversal are an intrinsically fourth-order feature in the spherical harmonic expansion.
They can best be modelled by introducing the fourth-rank tensor $\tsor{C}$
as defined in~\equ{she4} explicitly, and are not adequately captured if $\tsor{C}$ is approximated by a closure in terms of $\QQ$.

Given that closure of the hierarchy of tensors appearing in the spherical harmonic expansion of $P(\vec{p})$
cannot be achieved at the $\QQ$ level, one could be tempted  to develop instead evolution models directly for $P(\vec{p})$,
or even for the entire pair correlation function $g(\vec{r})$~\citep{nazockdast_microstructural_2012,nazockdast_pair-particle_2013}.
However these approaches also require closures
(three- and higher point correlations need to be expressed in terms of the two-point one),
which might be hard to establish in the absence of powerful results like those of~\citet{Hand1962} that
strongly limit the number of possibilities.

On the other hand we showed that $P(\vec{p})$ and its time derivative
$\dgam{P}(\vec{p})$ were both reasonably well approximated by their fourth-order spherical harmonic expansion.
This suggests the possibility of developing accurate closed evolution equations for $\QQ$ and $\tsor{C}$,
that is, closing the hierarchy at the fourth-rank $\tsor{C}$ level, effectively
expressing the residual higher order spherical harmonics
in terms of $\QQ$ and $\tsor{C}$.
One step on this route could be an extension of Hand's result to the fully symmetric fourth-rank tensor $\tsor{C}$.
However, one expects this to still lead to a high dimensional parameter space.

An alternative to postulating general forms and then fitting a large number of parameters
is to base a simplified description on a micromechanical `kinetic' theory, including only the terms suggested by that theory.
Indeed this has been a major element in fabric evolution models so far~\citep{Hinch1976,kuzuu_constitutive_1983,Phan-Thien1995}.
It is perhaps the best way to avoid the generic blowups that emerged above beyond linear order in $\QQ$
(since a judicious kinetic theory will presumably map onto parameters within a stable basin of attraction).
Clearly, though, in attempting to capture the orientational distribution of near-contacts,
this avenue is subject to the same difficulties as outlined above unless a fourth rank tensor is introduced.

To follow either of these paths to an improved model, further extensive simulations could prove useful.
In particular, in order to gain insight in the dynamical coupling of $\tsor{C}$ and $\QQ$ tensors,
it may be enlightening to simulate a purely extensional flow.
The additional symmetry beyond that of the shear reversal flows addressed here allows no possible misalignment of the fabric tensor $\QQ$
with the flow tensor $\hat{\tsor{E}}$, and any four-lobed structures are restricted to those symmetric about the flow axis.
In three dimensions, the reversal of a uniaxial extensional flow is a biaxial extensional flow, with therefore an inequivalent steady state.
This complicates matters, but as we have seen, there is much insight to gain from two-dimensional simulations and by extension
from three-dimensional planar extensional flow simulations.
In that case forward and backward extensional flow are equivalent with the same steady-state microstructure, up to a rotation.
Simulations of such extensional flows are achievable even for infinite extensional strains thanks
to specific periodic boundary setups~\citep{kraynik_extensional_1992,seto_microstructure_2017}.

It would also be interesting to generalize our studies of shear reversal to the case of shear rotation,
in which the direction of shearing is smoothly or suddenly rotated by an angle of less than 180 degrees.
Shear reversal is the extreme case where much of the contact network
is destroyed before being recreated with the opposite orientation,
and the challenges faced by fabric evolution models are most acute in the middle of this process.
Such models {\em might} be more successful in flows that nudge the fabric from one orientation
to another via modest or continuous changes of flow direction.
Since the rate-independence of dense suspension flows mean that all non-reversing shear flow histories with fixed axes are equivalent,
shear flows with a nontrivial flow-axis history are perhaps the closest analog of time-dependent flows in ordinary viscoelastic materials.

%% file: appendix_numerical_method.tex

\section{Numerical method}\label{sec:appendix_numerics}
The equation of motion for $N$ spheres without inertia is simply
the $6N$-dimensional force/torque balance between hydrodynamic ($\vec{F}_{\mathrm{H}}$)
and contact ($\vec{F}_{\mathrm{C}}$) interactions,
which depend on the positions $\vec{X}$ and the velocities/angular velocities $\vec{U}$,
\begin{equation}
  \vec{0} = \vec{F}_{\mathrm{H}}(\vec{X},\vec{U}) + \vec{F}_{\mathrm{C}}(\vec{X}).
  \label{eq:force_balance}
\end{equation}

Decomposing the flow $\vec{v}(\vec{r})=\vec{\omega}\times \vec{r} + \tsor{e}\cdot\vec{r}$ in rotational
$\vec{\omega} = (0,0,-\sg/2)$ and extensional $\tsor{e}_{12}=\tsor{e}_{21}=\sg/2$ parts,
the lubrication force and torque vector takes the form
\begin{equation}
  \vec{F}_{\mathrm{H}}(\vec{X},\vec{U}) =
  -\tsor{R}_{\mathrm{FU}}(\vec{X}) \cdot \bigl(\vec{U}-\vec{U}^{\mathrm{flow}} \bigr)
  + \tsor{R}_{\mathrm{FE}}(\vec{X}):\tsor{E}, \label{eq:hydro_force}
\end{equation}
with $\vec{U}^{\mathrm{flow}} = (\vec{v}(y_1), \dots, \vec{v}(y_N), \vec{\omega}(y_1), \dots, \vec{\omega}(y_N))$
and $\tsor{E} = (\tsor{e}(y_1), \dots, \tsor{e}(y_N))$.
The position-dependent resistance second-rank tensor
$\tsor{R}_{\mathrm{FU}}$ and third-rank $\tsor{R}_{\mathrm{FE}}$ include the
so-called ``squeeze'', ``shear'' and ``pump'' modes of lubrication~\citep{ball_simulation_1997}.
The occurrence of contacts between particles in actual suspensions (due to surface roughness, finite-slip boundary conditions, or other factors) is mimicked
by a regularization of the resistance divergence at vanishing interparticle gap
$h_{ij} = 2(r_{ij}-a_i-a_j)/(a_i+a_j)$: the ``squeeze'' mode is $\propto 1/(h+\delta)$ and
the ``shear'' and ``pump'' modes are $\propto \log(h+\delta)$~\citep{Mari2014},
with $\delta=10^{-3}$.

Contacts are modelled by a pair of linear springs and dashpots consisting of
both normal and tangential components, a simple model
commonly used in granular physics~\citep{cundall_discrete_1979}.
In order to stay as close as possible to a hard sphere behavior,
the spring stiffnesses are taken such that the largest particle overlaps never exceed \SI{2}{\percent}
of particles' radii during the simulation.
By doubling this value, we checked that the shear reversal
dynamics is sensitive to the value
of the allowed overlap up to
a strain of around \SI{1}{\percent} after reversal, and is independent of it for later strains.

Equation~\eq{eq:force_balance} is solved for the velocities,
from which the positions are updated at every time step
via a mid-point algorithm.

%% file: appendix_componentwise_hand.tex

\section{Componentwise Hand equation for simple shear flow\label{Handapp}}
\subsection{In three dimensions}

Recall \equ{eq:Handeq_general} and the invariants \hbox{$I_1$--$I_{10}$}.
Some of these invariants are trivial in our case.
The tensors $\QQ$ (by definition) and $\tsor{E}$ (due to incompressibility) are traceless, i.e. $I_1=I_4=0$.
Moreover, past the instant of shear reversal at $\gamma=0$, $\tsor{E}$ and thus its powers are constant,
which means that $I_5$ and $I_6$ can be absorbed into constant coefficients.
We also note that in our case of simple shear flow $\vec{v} = (\sg y,0,0)$ we have
(recalling $\tsor{\hat{E}}=\tsor{E}/\ag$)
\begin{align}
\tsor{\hat{E}}^2 \cdot \QQ + \QQ \cdot \tsor{\hat{E}}^2 &= \frac{1}{2} \QQ + 2 I_9 \left( \Id - 4 \tsor{\hat{E}}^2 \right) \\
\tsor{\hat{E}}^2 \cdot \QQ^2 + \QQ^2 \cdot \tsor{\hat{E}}^2 &= \frac{1}{2} \QQ^2 - 8 \left(I_9\right)^2 \left( \Id - 4 \tsor{\hat{E}}^2 \right),
\end{align}
so the parameters $ \alpha_7 $ and $ \alpha_8 $ can be set to zero without any loss of generality.

In addition, $Q_{13}$ and $Q_{23}$ vanish due to the symmetry with respect to the shear plane.
In consequence $\QQ$ only has three independent components $Q_{\pm} := Q_{11} \pm Q_{22}$ and $\xy{Q}$.
The Hand equation \eq{eq:Handeq_general} for these three components then yields
\begin{align}
\xmy{\dgam{Q}} &= \left[ \alpha_1 + \alpha_3 \xpy{Q}  \right] \sgg \xmy{Q} + 2 \xy{Q}, \label{Hand_dy1}\\
\xpy{\dgam{Q}} &= \frac{1}{6} \alpha_4 \sgg + \alpha_1 \sgg \xpy{Q} + \frac{2}{3} \alpha_5 \xy{Q}  \nonumber \\
 &+ \alpha_3 \sgg \left( \frac{1}{6} \xmy{Q}^2 - \frac{1}{2} \xpy{Q}^2 + \frac{2}{3} \xy{Q}^2 \right) + \frac{2}{3} \alpha_6 \xpy{Q} \xy{Q}, \label{Hand_dy2} \\
\xy{\dgam{Q}} &= \frac{1}{2} \alpha_2  - \frac{1}{2} \xmy{Q} + \frac{1}{2} \alpha_5 \xpy{Q} + \alpha_1 \sgg \xy{Q} \nonumber \\
&+ \alpha_6 \left( \frac{1}{4} \xmy{Q}^2 + \frac{1}{4} \xpy{Q}^2 + \xy{Q}^2 \right) + \alpha_3 \sgg \xpy{Q} \xy{Q}.  \label{Hand_dy3}
\end{align}

Furthermore, in our flow
\begin{align}
I_2 &= \frac{1}{2} \xmy{Q}^2 + \frac{3}{2} \xpy{Q}^2 + 2 \xy{Q}^2 \\
I_3 &= \frac{3}{4} \xmy{Q}^2 \xpy{Q} - \frac{3}{4} \xpy{Q}^3 + 3 \xpy{Q} \xy{Q}^2 \\
I_7 &= \sgg \xy{Q} \label{I7} \\
I_8 &= \sgg \xpy{Q} \xy{Q} \label{I8} \\
I_9 &= \frac{1}{4} \xpy{Q} \label{I9} \\
I_{10} &= \frac{1}{8} \xmy{Q}^2 + \frac{1}{8} \xpy{Q}^2 + \frac{1}{2} \xy{Q}^2 \label{I10}.
\end{align}
Since $ \sgg \xy{Q} = I_7 $, $ \xpy{Q} = 4 I_9 $ and $ \xmy{Q}^2 = 8 I_{10} - I_7^2 - 16 I_9^2 $, we see that $ I_2 $, $ I_3 $ and $ I_8 $ are polynomials in $ I_7 $, $ I_9 $ and $ I_{10} $, and that an analytic function in the invariants is really an analytic function in $ \sgg \xy{Q} $, $ \xpy{Q}$, and $\xmy{Q}^2$.
It is thus clear that \equ{eq:componentwise_3d} hold true.

To show that the Hand equation \eq{eq:Handeq_general} does not constrain $\xmy{\dgam{Q}}$, $\xpy{\dgam{Q}}$,
and $\xy{\dgam{Q}}$ any further than equations \equ{eq:componentwise_3d},
consider \equ{eq:Handeq_general} after enforcing tracelessness and proportionality to $\sg$:
\begin{multline}
\sgg \dgam{Q} = \tsor{\hat{W}} \cdot \QQ - \QQ \cdot \tsor{\hat{W}} + \beta_1 \QQ + 2 \left( \beta_2 - \beta_1 \hat{I}_7 \right) \tsor{\hat{E}}  \nonumber \\
 +6 \left( \beta_4 - 4 \beta_1 \hat{I}_9 \right) \tsor{\hat{E}}^2 - 2 \left( \beta_4 - 4 \beta_1 \hat{I}_9 \right) \hat{I}_5 \Id,
\end{multline}
corresponding for our flow to the coupled system
\begin{align}
\xmy{\dgam{Q}} &= \beta_1 \sgg \xmy{Q} + 2 \xy{Q}, \\
\xpy{\dgam{Q}} &= \beta_4 \sgg, \\
\xy{\dgam{Q}} &= \beta_2 - \frac{1}{2} \xmy{Q}.
\end{align}
Since, as we have established, the only constraint the Hand equation \eq{eq:Handeq_general} imposes on the tensor coefficients $\beta_i$ is that they are are analytic functions of $ \sgg \xy{Q} $, $ \xpy{Q}$ and $\xmy{Q}^2$,
we may conclude that \equ{eq:componentwise_3d} are the most general allowed in three dimensions.

\subsection{In two dimensions\label{hand2d}}
In two dimensions,
the tensorial second-order spherical harmonic expansion
(in this case a Fourier series expansion) of
a probability density $P(\vec{p})$ for
pairs with centre-to-centre orientation $\vec{p}$ is
\begin{equation}
P(\vec{p}) \approx \frac{1}{2 \pi} \left( 1 + 4 \QQ : \vec{p} \vec{p} \right),
\end{equation}
where
$\QQ \equiv \langle \vec{p}  \vec{p} \rangle - \frac{1}{2} \Id $.

We need to re-derive the Hand equation \eq{eq:Handeq_general} in two dimensions.
We start this by noting that the result from frame-indifference~\citep{Noll1955}
\begin{equation}
\dQ = \tsor{W} \cdot \QQ - \QQ \cdot \tsor{W} + \mathbb{F} \left( \QQ, \tsor{E} \right)
\end{equation}
holds in two dimensions as it does in three.
Put differently, the two
dimensional case can be seen as a special case of the
three-dimensional case with the axis of the rigid rotation
along the vorticity axis.

From equation (8.13) of \citet{Rivlin1955} (after applying equation (4.7)), any
polynomial in symmetric $2 \times 2$ tensors $\tsor{A} $
and $ \tsor{B} $ can be written in the form
\begin{equation}
\varphi_0 \Id + \varphi_1 \tsor{A} + \varphi_2 \tsor{B},
\end{equation}
where the $\varphi_i$ are polynomials in the invariants
(see paragraph in \citet{Rivlin1955} below Eq. 13.3)
$ \mathrm{Tr} \tsor{A}$, $\mathrm{Tr} \tsor{B}$,
$ \mathrm{Tr} \tsor{A}^2$, $ \mathrm{Tr} \tsor{B}^2$,
and $\mathrm{Tr} \tsor{A} \cdot \tsor{B}$.

We can therefore write the two-dimensional Hand equation for
$\QQ$,
\begin{equation}
\sgg \dgam{Q} = \tsor{\hat{W}} \cdot \QQ - \QQ \cdot \tsor{\hat{W}} + \alpha_1 \QQ + \alpha_2 \tsor{\hat{E}},
\end{equation}
or in component form,
\begin{equation}
  \begin{aligned}
  \xmy{\dgam{Q}} &= \alpha_1 \sgg \xmy{Q} + 2 \xy{Q} \\
  \xy{\dgam{Q}} &= \frac{1}{2} \alpha_2 - \frac{1}{2} \xmy{Q} + \alpha_1 \sgg \xy{Q}, \label{dy3_2d}\\
  \end{aligned}
\end{equation}
with the $\alpha_i$ analytic functions of the invariants
$ I_1 \equiv \mathrm{Tr} \left( \QQ \right) $,
$ I_2 \equiv \mathrm{Tr} \left( \QQ^2 \right) $,
$ I_4 \equiv \mathrm{Tr} \left( \tsor{\hat{E}} \right) $,
$ I_5 \equiv \mathrm{Tr} \left( \tsor{\hat{E}}^2 \right) $,
and
$ I_7 \equiv \mathrm{Tr} \left( \QQ \cdot \tsor{\hat{E}} \right) $
Here the labelling is chosen so as to be consistent with the
three-dimensional case.

For an incompressible shear flow in two dimensions $\vec{v} = \left(\sg y, 0 \right)$,
the invariants are
\begin{align}
I_1 &= 0, \\
I_2 &= \frac{1}{2} \xmy{Q}^2 + 2 \xy{Q}^2, \\
I_4 &= 0, \\
I_5 &= \frac{1}{2}, \\
I_7 &= \sgg \xy{Q}.
\end{align}
We see that an analytical function of the invariants is an
analytical function of $ \sgg \xy{Q} $
and $\xmy{Q}^2$, so that the most general form of the two-dimensional Hand equation for
our system is
\begin{align}
\xmy{\dgam{Q}} &= \xmy{P}\left[ \sgg \xy{Q}, \xmy{Q}^2 \right] \sgg \xmy{Q} + 2 \xy{Q}, \\
\xy{\dgam{Q}} &= \xy{P}\left[ \sgg \xy{Q}, \xmy{Q}^2 \right]
- \frac{1}{2} \xmy{Q},
\end{align}
where the $P_{ij}$ are analytical functions of their arguments.

%% file: appendix_Qdot_decomp.tex

\section{Decomposition of the fabric evolution\label{Q_decomp}}

The near-contact angular distribution at a strain $\gamma$ is defined as
\begin{equation}
    P_{\gamma}(\vec{p}) = \frac{1}{N_{\gamma}} \sum_i \delta\left[\vec{p} - \vec{p}_i(\gamma)\right].
\end{equation}
Between any two successive strain steps $\gamma$ and $\gamma+\mathrm{d}\gamma$,
we can separate the evolution of this distribution
into three kinds of events:
\begin{itemize}
\item a near-contact disappears between $\gamma$ and
$\gamma+\mathrm{d}\gamma$ (death),
\item a near-contact appears between $\gamma$ and
$\gamma+\mathrm{d}\gamma$ (birth),
\item a near-contact survives between $\gamma$ and
$\gamma+\mathrm{d}\gamma$ and is just advected.
\end{itemize}
Hence we can write
\begin{equation}
    \begin{split}
     P_{\gamma+\mathrm{d}\gamma}(\vec{p}) - P_{\gamma}(\vec{p}) =
        & \sum_{i \in \text{ advected}} \frac{1}{N_{\gamma+\mathrm{d}\gamma}} \delta\left[\vec{p} - \vec{p}_i(\gamma+\mathrm{d}\gamma)\right]
                                            - \frac{1}{N_{\gamma}} \delta\left[\vec{p} - \vec{p}_i(\gamma)\right] \\
        & + \frac{1}{N_{\gamma+\mathrm{d}\gamma}} \sum\limits_{i \in \text{ birth}} \delta\left[\vec{p} - \vec{p}_i(\gamma+\mathrm{d}\gamma)\right] \\
        & - \frac{1}{N_{\gamma}} \sum\limits_{i \in \text{ death}} \delta\left[\vec{p} - \vec{p}_i(\gamma)\right].
    \end{split}
\end{equation}

Now, calling $N_{\text{birth}}$ and $N_{\text{death}}$ the number of near-contacts respectively being born and dying between
$\gamma$ and $\gamma+\mathrm{d}\gamma$, we define
\begin{align}
    \dgam{P}^{\mathrm{a}}(\vec{p}) & := \frac{1}{\mathrm{d}\gamma}\sum_{i \in \text{ advected}} \frac{1}{N_{\gamma+\mathrm{d}\gamma}} \delta\left[\vec{p} - \vec{p}_i(\gamma+\mathrm{d}\gamma)\right]
                                              - \frac{1}{N_{\gamma}} \delta\left[\vec{p} - \vec{p}_i(\gamma)\right] \\
    P^{\mathrm{b}}(\vec{p}) & := \frac{1}{N_{\text{birth}}} \sum\limits_{i \in \text{ birth}} \delta\left[\vec{p} - \vec{p}_i(\gamma+\mathrm{d}\gamma)\right] \\
    P^{\mathrm{d}}(\vec{p}) & := \frac{1}{N_{\text{death}}} \sum\limits_{i \in \text{ death}} \delta\left[\vec{p} - \vec{p}_i(\gamma)\right]
\end{align}
so that
\begin{equation}
    \dgam{P}_{\gamma}(\vec{p}) = \dgam{P}^{\mathrm{a}}(\vec{p}) + r^{\mathrm{b}} P^{\mathrm{b}}(\vec{p}) - r^{\mathrm{d}} P^{\mathrm{d}}(\vec{p})
\end{equation}
which is \equ{eq:Pdot_decomposition} in the main text, with the birth and death rates
$r^{\mathrm{b}} = N_{\text{birth}}/(N_{\gamma+\mathrm{d}\gamma}\mathrm{d}\gamma)$ and $r^{\mathrm{d}} = N_{\text{death}}/(N_{\gamma}\mathrm{d}\gamma)$.

Taking the traceless second moment of $\dgam{P}^{\mathrm{a}}$, $P^{\mathrm{b}}$ and $P^{\mathrm{d}}$,
we can also get a decomposition of the strain derivative of the fabric tensor:
\begin{equation}
    \dgam{\QQ} = \dgam{\QQ}^{\mathrm{a}} + r^{\mathrm{b}} \QQ^{\mathrm{b}}(\vec{p}) - \QQ^{\mathrm{d}} P^{\mathrm{d}}(\vec{p})
\end{equation}
which is \equ{Qbd} in the main text.

%% file: appendix_pdot_components_plot.tex

\section{Plots of advective, birth and death components of $\dot{P}(\vec{p})$\label{sec:appendix_Pdot_comp_plot}}

\includegraphics[width=\textwidth]{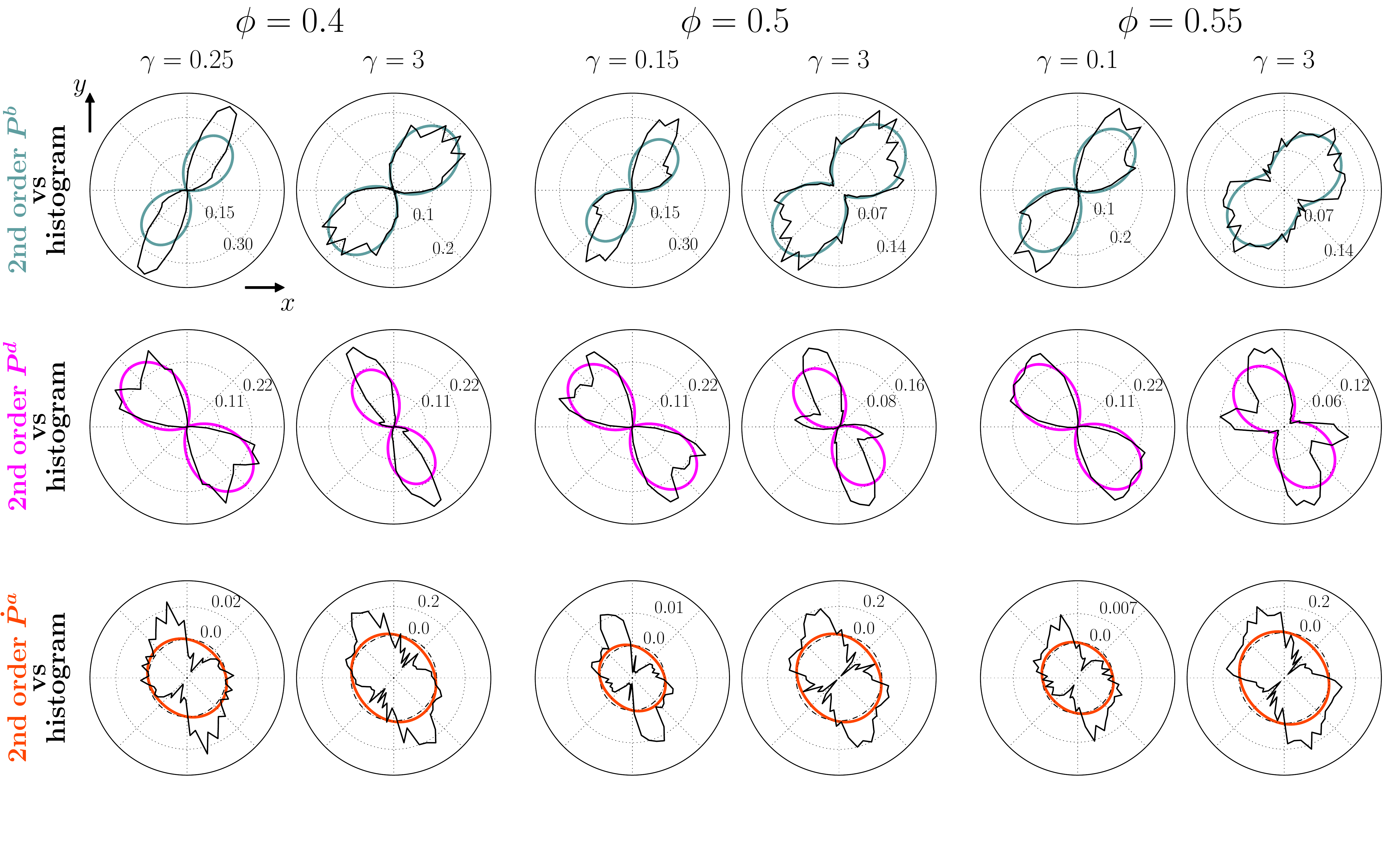}
Polar plots of $\dgam{P}^a(\vec{p})$ (\textbf{top}),
$P^b(\vec{p})$ (\textbf{middle}) and $P^d(\vec{p})$ (\textbf{bottom}),
as defined in \equ{eq:Pdot_decomposition}, in the shear plane (black lines) compared to
their second-order spherical harmonic approximations (color lines),
for $\phi=0.4$ (left column), $\phi=0.5$ (middle column), and $\phi=0.55$ (right column),
and for the two strain values after reversal indicated on top of each $\phi$ column.
Below are the same data compared to their fourth-order spherical harmonic approximations.

\includegraphics[width=\textwidth]{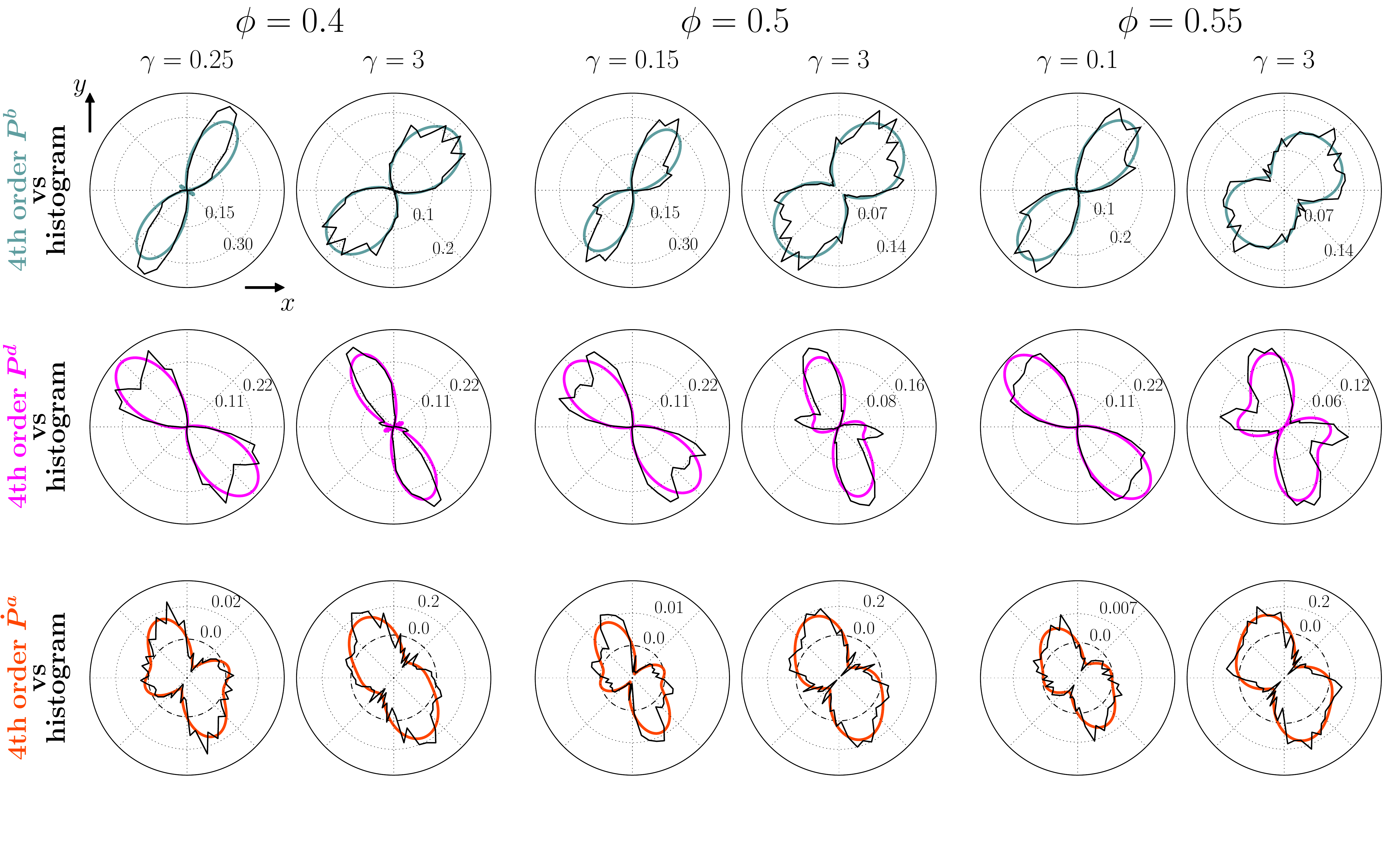}